\documentclass[aps,prc,twocolumn,superscriptaddress,amsmath,amssymb,%
floatfix]{revtex4-2}

\usepackage{graphicx}
\usepackage{bm}
\usepackage{booktabs}

\newcommand{\Ntot}{N_{\mathrm{tot}}}
\newcommand{\Scc}{\mathcal{S}_{cc}}
\newcommand{\om}{\vec{\omega}_g^{(c)}}
\newcommand{\sg}{\sigma_g^{(c)}}
\newcommand{\Fg}{\mathcal{F}_g^{(\tau)}}
\newcommand{\Fgi}{F_{g,i}^{(\tau)}}
\newcommand{\Res}{\operatorname{Res}}
\newcommand{\Tr}{\mathsf{T}}
\newcommand{\Fvec}{\vec{\mathcal{F}}^{(\tau)}}
\newcommand{\Fbg}{\vec{\mathcal{F}}^{(\tau)}_{\mathrm{bg}}}
\newcommand{\Fbgd}{\vec{\mathcal{F}}^{(\tau)\dagger}_{\mathrm{bg}}}
\newcommand{\cg}{c_g^{(\tau)}}
\newcommand{\qg}{q_g^{(\tau)}}
\newcommand{\pg}{p_g^{(\tau)}}
\newcommand{\omd}{\vec{\omega}_g^{(c)\dagger}}
\newcommand{\ReEg}{\operatorname{Re}E_g}

\begin{document}

\title{Fano formula for resonance poles in the Jost--random-phase
approximation:\\
Breit--Wigner-type peaks, Fano dips, and Fano-type peaks in $^{16}$O and
$^{208}$Pb}

\author{Kazuhito Mizuyama}
\email{mizukazu147@gmail.com}
\affiliation{Institute of Research and Development, Duy Tan University, Da Nang
550000, Vietnam}
\affiliation{Faculty of Natural Sciences, Duy Tan University, Da Nang 550000,
Vietnam}

\date{\today}

\begin{abstract}
Fano's formula is
derived for a resonance pole of the strength function in the continuum
random-phase approximation
formulated within the Jost function framework (Jost-RPA). At a pole, the residue
of the transition-amplitude vector of the open
channels lies along the Takagi vector of the pole for every external field. Of
the background, the amplitude vector without the pole term, only the component
along the Takagi vector therefore interferes with the resonance, and the
strength function can be written exactly in Fano's form for several continuum
channels. In this form $q$ is complex and energy dependent, and the modulus $p$
of the cosine of the angle between the Takagi vector and the background serves
as Fano's $|\rho|$. In an idealized form of Fano's model, $q$ and $p$ reduce to
Fano's $q$ and to the overlap $|\rho|$ of
Fano and Cooper. A second exact form separates a Fano term and a Breit--Wigner
term, the part of the resonance that does
not interfere with the background. Three patterns follow: Breit--Wigner-type
peaks for $p^2\simeq0$, Fano dips, and
Fano-type peaks for $p^2\simeq1$. A negative real part of the complex transition
probability, the residue of the response function at the pole, is equivalent to
Fano's condition $|q|<1$ for a dip at the resonance energy, evaluated at the
complex conjugate of the pole energy. Earlier
Jost-RPA analyses were made only for $^{16}$O. The analysis is applied to
$^{16}$O and to $^{208}$Pb, with up to 62 open
channels. It attributes the dips found earlier in $^{16}$O to the interference
of single poles with their backgrounds,
and it shows in both nuclei Fano dips, Fano-type peaks, and poles whose pattern
changes with the field. In $^{208}$Pb, narrow poles near the broad poles of the
giant dipole resonances produce Fano-type peaks with dips in the field of each
resonance. The patterns are not determined by the
collectivity indicators of the poles.
\end{abstract}

\maketitle

\section{Introduction}
\label{sec:intro}
Fano showed that a discrete state embedded in a continuum produces an asymmetric
line shape, the profile
$(\epsilon+q)^2/(1+\epsilon^2)$, whose parameter $q$ measures the excitation of
the discrete state relative to that of
the continuum~\cite{Fano1961}. When several continuum channels are present, the
discrete state interacts with only one
combination of them, and the orthogonal combinations add to the transition
probability without
interference~\cite{Fano1961,FanoCooper1965}. Line shapes of this kind, Fano
resonances, appear in many fields of physics~\cite{Miroshnichenko2010}, and a
part of the continuum that does not interfere has also been added to Fano's
profile in photonics~\cite{Limonov2017}. In near-threshold hadron scattering,
the interference between the pole term and a background term of the amplitude
can produce a dip in the cross section~\cite{SoneHyodo2025}. In nuclei, which
are open quantum systems coupled to the particle continuum~\cite{Okolowicz2003},
the interference of resonances through common decay channels has been studied
for giant resonances in the continuum shell model~\cite{Kleinwachter1986} and in
schematic models of
interfering doorway
states~\cite{Sokolov1997a,Sokolov1997b,AuerbachZelevinsky2011}. Asymmetric Fano
line shapes
have been discussed for core-excited resonances of exotic
nuclei~\cite{Orrigo2006}. Real, energy-dependent Fano
parameters were derived for neutron elastic scattering by open-shell nuclei in
the Hartree--Fock--Bogoliubov framework
with the two-potential formula, with the hole states of the Hartree--Fock
potential as the discrete
states~\cite{Mizuyama2020_FanoHFB}.

Decompositions of the $S$ matrix into a term singular at one resonance pole and
a background have been used to define
complex Fano parameters for the conductance of quantum
dots~\cite{RacecWulf2001}, energy dependent in
Ref.~\cite{Racec2010}. An isolated resonance is one well separated from the
other resonances compared with their
widths. For such a resonance coupled to more than two open channels, the Fano
parameter of a partial transmission
coefficient is in general complex, even in the presence of time-reversal
symmetry~\cite{Mendoza2008}. Fano's theory has been extended to overlapping
resonances interacting with several continuum channels~\cite{Mies1968}. Fano
parameters that depend on energy through the other resonances have been obtained
from the pole representation of the
$S$ matrix~\cite{Magunov2003}, and Fano parameters set by the interference with
other resonant, anti-resonant, and bound
states from resonant-state expansions~\cite{Sasada2011}; photoionization cross
sections have been expanded exactly over
resonances~\cite{Alvarez1989}. Fano resonances of extinction spectra have been
described with electromagnetic
quasinormal modes~\cite{Bochkarev2026}, and Fano's parameter in atomic
absorption spectra has been related to the phase
of the dipole response~\cite{Ott2013}. Residues of rank one, products of the
amplitudes for the decay of a resonance
into the channels, are standard in $R$-matrix theory~\cite{LaneThomas1958} and
in analyses of hadron resonances. In the latter, the factors are the pole
couplings to the channels, which are in general complex and enter the products
without complex conjugation~\cite{PDG2026_Resonances,Sekihara2015}. The
collision matrix~\cite{HumbletRosenfeld1961} and the $T$ matrix of
electromagnetic scatterers, with matrix-valued
residues~\cite{Fischbach2026}, have been expanded over their poles.

The continuum random-phase approximation (RPA)~\cite{ShlomoBertsch1975}
formulated within the Jost function framework (Jost-RPA) obtains the
open-channel $S$ matrix and its poles on the
Riemann sheets directly from the Jost functions of the coupled radial equations.
Reference~\cite{Mizuyama2023_JRPA1}
obtained the complex eigenenergies of the giant dipole resonance of $^{16}$O as
poles and showed that the strength
function agrees with that of the conventional continuum RPA. The strength
function was then written as the squared norm of an amplitude vector in the open
channels~\cite{Mizuyama2024_Eigenphase}, and the response function was
expanded over its poles by the Mittag-Leffler theorem~\cite{Mizuyama2025_JRPA2}.
In the quadrupole strength of
$^{16}$O, Ref.~\cite{Mizuyama2025_JRPA2} found a dip near 28.25~MeV in the
isoscalar field, where poles have residues
with negative real parts, and a dip in the isovector field near the peak energy
of the isoscalar giant quadrupole
resonance (ISGQR). It presented these dips as examples that are difficult to
explain without the Fano effect. The
Takagi factorization of the rank-one residue of the $S$ matrix was introduced
for the resonant states of the
Hartree--Fock--Bogoliubov theory in Ref.~\cite{Mizuyama2026_Takagi} and applied
within the Jost-RPA in
Ref.~\cite{Mizuyama2026_Collectivity}. The latter defined the transition
amplitudes of a pole for each configuration,
collectivity indicators built from them, and the branching ratios over the open
channels. It also related the isoscalar dip near 28.25~MeV to a pole with a
negative real part of the complex transition probability, the residue of the
response function at the pole. By matching
Fano's profile to the single-pole term of the strength function, it expressed
Fano's parameter as $q=-\cot\Theta_g$
through the phase $\Theta_g$ of the transition amplitude of the pole. All these
Jost-RPA analyses were made for
$^{16}$O.

These results leave several questions. Fano's formulation divides the system
into a discrete state and a continuum.
Can Fano's formula for several continuum channels be obtained from a pole of the
Jost-RPA strength function itself, without such
a division? What property of a pole and its background decides whether the pole
produces a peak or a dip, and whether
a peak comes from interference with the background? Why can the same pole
produce different line shapes in the
isoscalar and the isovector fields? Do the collectivity indicators determine the
line shape? What do a negative real
part of the complex transition probability and $q=-\cot\Theta_g$ mean in terms
of Fano's parameter? And do the answers
hold in a heavy nucleus, in which many channels are open?

The aim is a qualitative
understanding of how a single pole shapes the strength function. In this paper,
the term of one pole is separated from the transition-amplitude vector of the
open channels in the
Jost-RPA. Its residue lies along the Takagi vector of the pole for every
external field, so that the Takagi vector is
the combination of channels that interacts with the resonance, and the pole of
the $S$ matrix takes the place of Fano's discrete state. The strength function
then has exactly Fano's form for several continuum channels, obtained here
without dividing the system into a discrete state and a continuum, so that
resonances are not classified by their origin. Unlike the decompositions of
Refs.~\cite{RacecWulf2001,Racec2010,Mendoza2008}, the separation applies to the
amplitude vector of an external field, not to the $S$ matrix or a transmission,
and the pole term is the exact principal part of the Laurent expansion, with the
full energy dependence kept in the background;
Refs.~\cite{RacecWulf2001,Mendoza2008} take the background as constant near the
resonance, and Ref.~\cite{Racec2010} separates a resonant term with an
energy-dependent denominator. In an idealized form of Fano's model, the
parameters of the form obtained here reduce to Fano's
$q$ and to the overlap $|\rho|$ of Fano and Cooper
(Sec.~\ref{subsec:fanoformula}). The same strength is also written as a Fano
term plus a Breit--Wigner term, the part of the resonance that does not
interfere. From the two forms I define three patterns of the strength near a
pole, the Breit--Wigner-type peak, the Fano dip, and the Fano-type peak, and
show what bounds $|\rho|$. A negative real part of the complex transition
probability is shown to be Fano's condition $|q|<1$ for a dip at the resonance
energy, evaluated at the complex conjugate of the pole energy; $-\cot\Theta_g$
is a special case of Fano's parameter at the pole (Sec.~\ref{subsec:phase}). The
analysis is applied to the $2^+$ and $1^-$ excitations of $^{16}$O
and to the $1^-$ excitations of $^{208}$Pb, which have 103 channels, up to 62 of
them open on the sheets analyzed. In both nuclei Fano dips and Fano-type peaks
appear, and the pattern of a pole can change with the field.

The paper is organized as follows. Section~\ref{sec:formalism} summarizes the
strength function and the pole
quantities of the Jost-RPA used in the following sections.
Section~\ref{sec:fano} separates the pole term, derives the
two forms of the strength function, names the three patterns, and relates Fano's
parameter at the pole to the phase of
the transition amplitude. Sections~\ref{sec:O16} and \ref{sec:Pb} analyze poles
of $^{16}$O and $^{208}$Pb, and Sec.~\ref{sec:common} the dependence of the
patterns on the external field and their relation to the collectivity indicators
and to the phase at the pole. Section~\ref{sec:summary} summarizes the results.
Appendix~\ref{app:jrpa} gives the model, the numerical details, and the
derivations of the Jost-RPA quantities, and
Appendix~\ref{app:phase} derives the relation between the complex transition
probability and Fano's parameter at the
pole.

\section{Strength function and pole quantities in the Jost-RPA}
\label{sec:formalism}
This section summarizes the definitions and results of
Refs.~\cite{Mizuyama2023_JRPA1,Mizuyama2024_Eigenphase,Mizuyama2025_JRPA2,%
Mizuyama2026_Takagi,Mizuyama2026_Collectivity}
used in the following sections. The model, the Jost functions,
the open-channel $S$ matrix, the derivations, and the normalization convention
are given in
Appendix~\ref{app:jrpa}, so that the paper can be read on its own.

The excitations of multipolarity $L$ of a doubly closed-shell nucleus are
described by particle-hole (p-h)
configurations on a Woods--Saxon mean field with a residual interaction of
Skyrme type. A channel $i$ specifies the
nucleon species $q\in\{n,p\}$, the hole orbit $h_i$ with energy
$\epsilon_{h_i}<0$, and the angular momenta
$(l_i,j_i)$ of the particle coupled with the hole to $L$. There are $\Ntot$
channels: $\Ntot=16$ and 14 for the $2^+$
and $1^-$ excitations of $^{16}$O and $\Ntot=103$ for the $1^-$ excitations of
$^{208}$Pb. The thresholds $-\epsilon_h$ divide
the real energy axis into intervals. On the $n$th interval the forward channels
(Appendix~\ref{subsec:model}) of the $n-1$ least bound hole orbits
are open. The resonance poles $E_g=\ReEg-i\Gamma_g/2$ that shape the strength
there lie on the Riemann sheet
reached by continuing $E$ from the interval into the lower half plane, labeled
$\mathrm{RS}=n$ as in
Ref.~\cite{Mizuyama2026_Collectivity}. The upper half of every sheet coincides
with the physical sheet. The open-channel quantities of a pole on
$\mathrm{RS}=n$ refer to the $n_c$ channels $c$ open on the $n$th interval.

\subsection{Strength function as a squared norm}
\label{subsec:SFnorm}
For the isoscalar ($\tau=0$) or isovector ($\tau=1$) external field $f^{(\tau)}$
(Appendix~\ref{subsec:strength}), the strength function
$S_F^{(\tau)}=-\operatorname{Im}R_F^{(\tau)}/\pi$, with the RPA response
function $R_F^{(\tau)}$ of
Eq.~(\ref{eq:RF}), is the squared norm of a transition-amplitude vector in the
open
channels~\cite{Mizuyama2024_Eigenphase},
\begin{equation}
  S_F^{(\tau)}(E)=\vec{\mathcal{F}}^{(\tau)\dagger}(E^*)\,\vec{\mathcal{F}}^{(%
\tau)}(E),
  \label{eq:SF-norm}
\end{equation}
\begin{align}
  \vec{\mathcal{F}}^{(\tau)}(E)&\equiv\Scc(E)\sum_q\int_0^\infty\!dr\,\vec{%
\rho}^{\,(-)}_q(r;E)\,f_q^{(\tau)}(r)
  \nonumber\\
  &=\sqrt{\frac{2m}{\pi\hbar^2}}\sum_q\int_0^\infty\!dr\,\Psi^{(+)\Tr}_{F,c}%
(r;E)\,\xi_q(r)\,f_q^{(\tau)}(r).
  \label{eq:Fvec}
\end{align}
Here $\Scc$, the open-channel block of the flux-adjusted $S$
matrix~(\ref{eq:Smat}), is complex symmetric and unitary on the real interval;
$\Psi^{(+)}_{F,c}(r;E)$ are the open-channel scattering solutions of
Eq.~(\ref{eq:Psi}), and $\Psi^{(-)}_{F,c}(r;E)\equiv\Psi^{(+)*}_{F,c}(r;E^*)$.
The vectors
$\vec{\rho}^{\,(-)}_q(r;E)\equiv\sqrt{2m/(\pi\hbar^2)}\,\Psi^{(-)\Tr}_{F,c}%
(r;E)\,\xi_q(r)$ contain the hole vectors
$\xi_q$ (Appendix~\ref{app:jrpa}). The two lines of Eq.~(\ref{eq:Fvec}) are
related by
$\Psi^{(+)\Tr}_{F,c}=\Scc\Psi^{(-)\Tr}_{F,c}$. On the real axis
$\vec{\mathcal{F}}^{(\tau)\dagger}(E^*)=[\Fvec(E)]^\dagger$, written
$\vec{\mathcal{F}}^{(\tau)\dagger}(E)$ below, so
that $S_F^{(\tau)}=\|\Fvec\|^2$. By the second line of Eq.~(\ref{eq:Fvec}) and
Eq.~(\ref{eq:Psi}), $\Fvec(E)$ has poles
only at the poles of $\mathcal{S}$; its residue at a pole is given in
Sec.~\ref{sec:fano}.

\subsection{Residue and transition amplitudes of a pole}
\label{subsec:poleamp}
At a non-degenerate pole $E_g=\ReEg-i\Gamma_g/2$, a simple zero of
$\det\mathcal{J}^{(+)}(E)$
(Appendix~\ref{subsec:Smatrix}), the residue of $\Scc$ is complex symmetric and
of rank one. Its
Autonne--Takagi factorization~\cite{Autonne1915,Takagi1924} therefore has a
single nonzero singular value
$\sg$~\cite{Mizuyama2026_Takagi},
\begin{equation}
  \Res\bigl[\Scc,E_g\bigr]=\sg\,\om\,\vec{\omega}_g^{(c)\Tr},\qquad\|\om\|=1,
  \label{eq:takagi}
\end{equation}
with the normalized open-channel Takagi vector
$\om$~\cite{Mizuyama2026_Collectivity}. The
residue of the Green function at the pole factorizes as
$\vec{\psi}_g(r)\vec{\psi}_g^{\Tr}(r')$ [Eq.~(\ref{eq:resG})], with the wave
function of the pole
\begin{equation}
  \vec{\psi}_g(r)\equiv\sqrt{\frac{4m\sg}{i\hbar^2}}\;\Psi^{(+)*}_{F,c}%
(r;E_g^*)\,\om.
  \label{eq:psig}
\end{equation}
Its forward and backward parts,
$\vec{\psi}_g=(\vec{\psi}_{g,X},\vec{\psi}_{g,Y})^{\Tr}$, define the transition
density of the nucleons $q$ at the pole and its channel components,
\begin{align}
  \rho^{(g)}_q(r)&\equiv\vec{\psi}_g^{\Tr}(r)\,\xi_q(r)
  =\sum_{i\in
q}\bigl(\vec{\psi}_{g,X}(r)+\vec{\psi}_{g,Y}(r)\bigr)_i\,\tilde{\varphi}_{q,i}%
(r)
  \nonumber\\
  &\equiv\sum_{i\in q}\rho^{(g)}_{q,i}(r),
  \label{eq:trd}
\end{align}
with $\tilde{\varphi}_{q,i}$ of Eq.~(\ref{eq:phitilde}). The residue of the
response function then factorizes into the square of the transition amplitude of
the pole,
\begin{equation}
  \widetilde{B}_g^{(\tau)}\equiv\Res\bigl[R_F^{(\tau)},E_g\bigr]=\mathcal{F}%
_g^{(\tau)2},
  \label{eq:Btilde}
\end{equation}
\begin{align}
  \Fg&\equiv\sum_q\int_0^\infty\!dr\,f_q^{(\tau)}(r)\,\rho^{(g)}_q(r)=%
\sum_{i=1}^{\Ntot}\Fgi,
  \nonumber\\
  \Fgi&\equiv\int_0^\infty\!dr\,f_{q_i}^{(\tau)}(r)\,\rho^{(g)}_{q_i,i}(r),
  \label{eq:Famp}
\end{align}
where $\Fgi$ is the amplitude with which the external field excites the pole
through the p-h channel $i$. With the
phase $\Theta_g^{(\tau)}$ of the transition amplitude, the complex transition
probability, which tends to the real transition probability in the
narrow-resonance limit~\cite{Mizuyama2026_Collectivity}, is
\begin{equation}
  \widetilde{B}_g^{(\tau)}=|\Fg|^2e^{2i\Theta_g^{(\tau)}},\qquad
\Fg=|\Fg|e^{i\Theta_g^{(\tau)}};
  \label{eq:Theta}
\end{equation}
since the sign of the Takagi vector is a convention and $\Fg$ changes sign with
it [Eqs.~(\ref{eq:psig}) and (\ref{eq:Famp})], $\Theta_g^{(\tau)}$ is defined
modulo $\pi$. Only $e^{2i\Theta_g^{(\tau)}}$ enters the
results, and the tables give $\Theta_g^{(\tau)}$ between $-\pi$ and $\pi$ for
the sign of $\om$ returned by the numerical factorization. Like the coupling
squared at a resonance pole, which enters the complex compositeness of hadron
resonances~\cite{KinugawaHyodo2025}, $\widetilde{B}_g^{(\tau)}$ is the square,
not the squared modulus, of a complex amplitude.

\subsection{Collectivity indicators}
\label{subsec:collect}
A collective state is a coherent superposition of many p-h
excitations~\cite{Co2013}. For a resonance pole,
the alignment of the phases of the channel amplitudes is measured by the
coherence index
\begin{equation}
  C_g^{(\tau)2}\equiv\frac{|\Fg|^2}{\Bigl(\sum_{i=1}^{\Ntot}|\Fgi|\Bigr)^2},%
\qquad 0\le C_g^{(\tau)2}\le1,
  \label{eq:C}
\end{equation}
and the number of channels that share the amplitude by the participation rate
\begin{equation}
  \eta_g^{(\tau)}\equiv\frac{\Bigl(\sum_{i=1}^{\Ntot}|\Fgi|\Bigr)^2}{%
\Ntot\sum_{i=1}^{\Ntot}|\Fgi|^2},
  \qquad \frac{1}{\Ntot}\le\eta_g^{(\tau)}\le1,
  \label{eq:eta}
\end{equation}
where $\Ntot\eta_g^{(\tau)}$ is an effective number of participating
channels~\cite{Mizuyama2026_Collectivity}. This number is the squared ratio of
the $\ell_1$ and $\ell_2$ norms of
$(F^{(\tau)}_{g,1},\dots,F^{(\tau)}_{g,\Ntot})$, a ratio used as a measure of
sparseness~\cite{Hoyer2004,Hurley2009}.
The product of $C_g^{(\tau)2}$ and $\Ntot\eta_g^{(\tau)}$,
\begin{equation}
  C_g^{(\tau)2}\Ntot\eta_g^{(\tau)}=\frac{|\Fg|^2}{\sum_{i=1}^{\Ntot}|\Fgi|^2}%
\le\Ntot,
  \label{eq:CNeta}
\end{equation}
equals one when a single channel carries the amplitude, is much larger than one
for a giant collective resonance, in
which many channels add in phase, and is much smaller than one when many
channels contribute with phases that
cancel~\cite{Mizuyama2026_Collectivity}. Below, this product is called the
indicator of the pole in the field $\tau$. These indicators
depend only on the ratios of the channel amplitudes. With Eq.~(\ref{eq:Theta}),
the product factorizes the complex
transition probability,
\begin{equation}
  \widetilde{B}_g^{(\tau)}=\Bigl[C_g^{(\tau)2}\Ntot\eta_g^{(\tau)}\Bigr]
  \Bigl(\sum_{i=1}^{\Ntot}|\Fgi|^2\Bigr)e^{2i\Theta_g^{(\tau)}},
  \label{eq:Bfact}
\end{equation}
into a coherent enhancement, a single-particle scale, and a phase. The sum
$\sum_i|\Fgi|$ in Eqs.~(\ref{eq:C}) and
(\ref{eq:eta}) cancels in the product, so that the factorization is exact.

\subsection{Branching ratios and the residue width}
\label{subsec:brsum}
The singular value of Eq.~(\ref{eq:takagi}) equals the norm of the residue,
$\sg=\|\Res[\Scc,E_g]\|$, and is
called the residue width. The components of $\om$ give the branching ratios over
the open channels~\cite{Mizuyama2026_Collectivity},
\begin{equation}
    \mathrm{br}_c\equiv|\omega^{(c)}_{g,c}|^2,\qquad \sum_{c}\mathrm{br}_c=1.
  \label{eq:br}
\end{equation}
The branching ratios are properties of the pole: they do not depend on the
external field. The residue width differs
in general from the pole width $\Gamma_g$; their ratio is
$\sg/\Gamma_g=\|\mathcal{S}_{\mathrm{B},g}(E_g^*)\,
\vec{\omega}_g^{(c)*}\|$ [Eq.~(\ref{eq:sigma-gamma})], where
$\mathcal{S}_{\mathrm{B},g}$ is $\Scc$ without the pole
term of $g$, so that $\sg=\Gamma_g$ when $\mathcal{S}_{\mathrm{B},g}$ does not
vary across the pole [Eq.~(\ref{eq:sigma-gamma-first})]. In the limit of a
narrow pole that does not overlap other poles, $\sg$ and $\mathrm{br}_c$ become
the decay width and the branching ratios of the
decay~\cite{Mizuyama2026_Collectivity}. For broad or overlapping hadron
resonances, which lie outside this limit, and for those near thresholds,
different definitions of the branching ratios can give rather different
values~\cite{Burkert2023}.

\section{Fano formula from the pole term}
\label{sec:fano}
In this section the term of a single pole $g$ is separated from the
amplitude~(\ref{eq:Fvec}). Throughout, $E$ lies on
the real energy interval of the sheet of the pole $g$ or on its continuation
into the complex plane, and quantities that
depend on the external field carry the label $(\tau)$. On this interval the
strength function~(\ref{eq:SF-norm}) is
rewritten exactly in two forms. The first is the form that Fano gave for a
discrete state coupled to several continuum
channels~\cite{Fano1961,FanoCooper1965}, and the second separates the part of
the resonance that does not interfere
with the background.

\subsection{Pole term of the transition-amplitude vector}
\label{subsec:poleterm}
In the first line of Eq.~(\ref{eq:Fvec}), the vectors
$\vec{\rho}^{\,(-)}_q(r;E)$ are analytic at $E=E_g$: the poles
of $\Psi^{(-)}_{F,c}(r;E)=\Psi^{(+)*}_{F,c}(r;E^*)$ lie at the complex
conjugates of the poles of $\mathcal{S}$, and its other
singularities are the threshold branch points on the real axis. The residue of
$\Fvec$ at $E_g$ therefore comes from
$\Scc$ alone, and Eq.~(\ref{eq:takagi}) gives
\begin{align}
  \Res\bigl[\Fvec,E_g\bigr]={}&\sg\,\om
  \nonumber\\
  &\times\sum_q\int_0^\infty\!dr\,\vec{\omega}_g^{(c)\Tr}\vec{\rho}^{\,(-)}%
_q(r;E_g)\,f_q^{(\tau)}(r).
  \label{eq:resF-1}
\end{align}
By the definition of $\vec{\rho}^{\,(-)}_q$ and Eqs.~(\ref{eq:psig}) and
(\ref{eq:trd}), the scalar under the
integral is proportional to the transition density of the pole,
\begin{align}
  \vec{\omega}_g^{(c)\Tr}\vec{\rho}^{\,(-)}_q(r;E_g)
  &=\sqrt{\frac{2m}{\pi\hbar^2}}\,\bigl[\Psi^{(+)*}_{F,c}(r;E_g^*)\,\om\bigr]^{%
\Tr}\xi_q(r)
  \nonumber\\
  &=\sqrt{\frac{i}{2\pi\sg}}\;\rho^{(g)}_q(r),
  \label{eq:omrho}
\end{align}
and Eq.~(\ref{eq:Famp}) gives
\begin{equation}
  \Res\bigl[\Fvec,E_g\bigr]=\cg\,\om,\qquad
  \cg\equiv\sqrt{\frac{i\sg}{2\pi}}\;\Fg,
  \label{eq:resF}
\end{equation}
with the branch of the square root used in Eq.~(\ref{eq:psig}). The residue
points along the Takagi vector $\om$
for every external field; the field enters only through the complex coefficient
$\cg$, that is, through the transition amplitude $\Fg$ of the pole. Since $\Fg$
changes sign together with $\om$, the product $\cg\om$ does not
depend on the sign convention of the Takagi vector. The residue of a
pion-photoproduction helicity multipole at the pole of a baryon resonance has
the same form: up to spin, isospin, and kinematic factors, it is the complex
photo-decay amplitude at the pole times the square root of the elastic $\pi N$
residue~\cite{Workman2013}.

Separating this pole term,
\begin{equation}
  \Fvec(E)=\Fbg(E)+\frac{\cg\,\om}{E-E_g},
  \label{eq:Fsplit}
\end{equation}
defines the background vector $\Fbg(E)$ of the pole $g$; $\Fbg$ and the
quantities built from it below refer to this
pole. The pole term is the principal part of the Laurent expansion of $\Fvec$ at
$E_g$, that is, the term of the pole
$g$ in a Mittag-Leffler expansion over the poles, which
Ref.~\cite{Mizuyama2025_JRPA2} applied to the response
function. For the amplitude defined with the Jost solutions of
Eq.~(\ref{eq:Psi}), whose asymptotic normalization
also defines $\mathcal{S}$ and its residue, the separation is therefore unique
and involves no approximation. The background is analytic at $E_g$ and carries
all
other singularities of $\Fvec(E)$: the other poles of $\Scc$ and the threshold
branch points. Only the term of $g$ is
separated, so that neither the expansion over all poles nor the conditions for
its convergence are needed.
\subsection{Fano's profile}
\label{subsec:qdef}
Fano~\cite{Fano1961} considered a discrete state $\varphi$ coupled to a
continuum of states $\psi_E$ by the matrix
element $V_E$ and excited from an initial state $|0\rangle$ by a transition
operator $T$ (Fano's notation, not a
$T$ matrix). The final state with the incoming-wave boundary condition is
$\Psi^{(-)}_E=-e^{-i\Delta}\Psi_E$, where $\Psi_E$ is Fano's standing-wave
eigenstate and the phase is chosen such that
$\Psi^{(-)}_E$ tends to $\psi_E$ away from the resonance. At each real $E$, the
amplitude to this final state is
\begin{equation}
  \langle\Psi^{(-)}_E|T|0\rangle=\langle\psi_E|T|0\rangle\,\frac{\bar{\epsilon}%
(E)+q(E)}{\bar{\epsilon}(E)+i},
  \label{eq:fano-amp}
\end{equation}
where $\bar{\epsilon}=-\cot\Delta$ is given by the phase shift $\Delta(E)$
caused by the coupling, and $q(E)$ is
Fano's parameter, not to be confused with the nucleon label $q$ of
Sec.~\ref{sec:formalism}. This parameter,
$q(E)=\langle\Phi_E|T|0\rangle/[\pi V_E^*\langle\psi_E|T|0\rangle]$ with the
discrete state
$\Phi_E$ modified by an admixture of the continuum~\cite{Fano1961}, measures the
excitation of the discrete state
relative to that of the continuum. The
squared modulus of Eq.~(\ref{eq:fano-amp}) divided by
$|\langle\psi_E|T|0\rangle|^2$ is Fano's profile
$|\bar{\epsilon}+q|^2/(1+\bar{\epsilon}^2)$~\cite{Fano1961,Miroshnichenko2010}.
In Fano's idealization, $V_E$, the
matrix elements of $T$, and the level shift do not depend on $E$. Then
$\bar{\epsilon}=(E-E_r)/(\Gamma/2)$ with a fixed
resonance energy $E_r$ and width $\Gamma=2\pi|V_E|^2$, and the resonant part
$\langle\Psi^{(-)}_E|T|0\rangle-\langle\psi_E|T|0\rangle=(\Gamma/2)(q-i)%
\langle\psi_E|T|0\rangle/(E-E_r+i\Gamma/2)$
is a pure pole term.

Since $\Psi^{(+)\Tr}_{F,c}(r;E)=\Psi^{(-)\dagger}_{F,c}(r;E)$ on the real axis,
the second line of
Eq.~(\ref{eq:Fvec}) shows that $\Fvec(E)$ is the amplitude to the open-channel
scattering states with incoming-wave
boundary conditions, the counterpart of $\langle\Psi^{(-)}_E|T|0\rangle$. Fano's
amplitude to the standing-wave state
$\Psi_E$, in contrast, has no simple pole. In place of $\bar{\epsilon}$, the
energy is measured from the pole $E_g$,
\begin{equation}
  \epsilon\equiv\frac{E-\ReEg}{\Gamma_g/2},\qquad
E-E_g=\frac{\Gamma_g}{2}\,(\epsilon+i).
  \label{eq:eps}
\end{equation}

\subsection{Fano formula for several open channels}
\label{subsec:fanoformula}
For a discrete state coupled to several continuum channels, Fano~\cite{Fano1961}
observed that the state interacts with only one combination of them; the
orthogonal combinations contribute to the transition probability without
interference. The pole term~(\ref{eq:Fsplit}) has this structure: its residue
$\cg\om$ points along the Takagi vector
$\om$ for every external field, so that $\om$ is the combination of open
channels that interacts with the resonance.
Splitting the background into its component along $\om$ and the orthogonal rest,
Eq.~(\ref{eq:Fsplit}) becomes
\begin{align}
  \Fvec(E)={}&\Bigl[\omd\Fbg(E)+\frac{\cg}{E-E_g}\Bigr]\om
  \nonumber\\
  &+\bigl(\mathbf{1}-\om\omd\bigr)\Fbg(E).
  \label{eq:Fomega}
\end{align}
The orthogonal rest has no pole term [Fig.~\ref{fig:schematic}(a)]. Along $\om$
the amplitude is that of one continuum, and with
Eq.~(\ref{eq:eps}) it takes Fano's form~(\ref{eq:fano-amp}),
\begin{align}
  \omd\Fbg(E)+\frac{\cg}{E-E_g}&=\omd\Fbg(E)\,\frac{\epsilon+\qg(E)}{%
\epsilon+i},
  \nonumber\\
  \qg(E)-i&\equiv\frac{\cg}{(\Gamma_g/2)\,\omd\Fbg(E)},
  \label{eq:qg}
\end{align}
at every real $E$ where $\omd\Fbg(E)\neq0$. The parameter $\qg$ plays the role
of Fano's $q$: $|\qg-i|$ is the ratio of
the amplitude of the pole term at the center, $|\cg|/(\Gamma_g/2)$, to the
background amplitude along $\om$. Outside Fano's idealization Fano's resonant
part is in general not a pure pole term, and Eq.~(\ref{eq:qg}) is a
definition, chosen so that the form of Eq.~(\ref{eq:fano-amp}) holds exactly,
not a result derived from Fano's theory.
Because, unlike Fano's theory and Feshbach's projection
formalism~\cite{Feshbach1958,Feshbach1962}, it requires no
division of the system or of its Hamiltonian into a discrete state and a
continuum, it does not classify resonances by
their origin.

With
\begin{equation}
  \pg(E)\equiv\frac{|\omd\Fbg(E)|}{\|\Fbg(E)\|},\qquad 0\le\pg(E)\le1,
  \label{eq:pg}
\end{equation}
the modulus of the cosine of the angle between $\Fbg(E)$ and $\om$ in the space
of the open channels, the squared norm
of Eq.~(\ref{eq:Fomega}) is
\begin{align}
  S_F^{(\tau)}(E)={}&p_g^{(\tau)2}(E)\|\Fbg(E)\|^2\,\frac{|\epsilon+\qg(E)|^2}%
{1+\epsilon^2}
  \nonumber\\
  &+\bigl[1-p_g^{(\tau)2}(E)\bigr]\|\Fbg(E)\|^2.
  \label{eq:fano}
\end{align}
Equation~(\ref{eq:fano}) has the form of Fano's result for a discrete state
coupled to several continuum channels~\cite{Fano1961,FanoCooper1965}. The first
term,
$\sigma_a(\epsilon+q)^2/(1+\epsilon^2)$ in the notation of
Ref.~\cite{FanoCooper1965} with $\sigma_a=p_g^{(\tau)2}\|\Fbg\|^2$, is the
transition
probability into the combination of continuum channels that interacts with the
resonance, and the second,
$\sigma_b=[1-p_g^{(\tau)2}]\|\Fbg\|^2$, that into the orthogonal combinations.
The two terms are called the
interfering term and the non-interfering background; Fig.~\ref{fig:schematic}(a)
shows which part of the background
gives each term. The ratio $p_g^{(\tau)2}$ is Fano's
$\rho^2=\sigma_a/(\sigma_a+\sigma_b)$, a ratio that has been
determined experimentally for Fano resonances in the predissociation of H$_2$
and D$_2$ from the energy dependence of
the angular distribution of the photofragments~\cite{WangMo2020}. In Eqs.~(2.13)
and (2.14) of
Ref.~\cite{FanoCooper1965}, Fano and Cooper identified $\rho$ as the overlap of
the direct dipole state (the continuum
state excited directly from the ground state) with the continuum state that
interacts with the discrete state.
The quantity $\pg$ is the modulus of the same overlap, with $\Fbg/\|\Fbg\|$ and
$\om$ in the places of these two states. Channel by
channel, Eq.~(\ref{eq:Fomega}) has the form of the photoelectron amplitude of
Kabachnik and Sazhina, Eq.~(45) of
Ref.~\cite{KabachnikSazhina1976}, with $\qg$ in place of Fano's $q$, and its
squared norm, Eq.~(\ref{eq:fano}),
corresponds to their Eqs.~(35) and (46). Here the Takagi
factorization~(\ref{eq:takagi}) supplies the
interacting combination of continuum channels from the pole itself, and no
assumption on the energy dependence of $\Fbg(E)$ is
made. Since $\cg$, $\om$, and $E_g$ are constants, $\qg$
and $\pg$ depend on $E$ only through the background; the field enters through
$\Fg$ and through the background, so that $\qg$ and $\pg$ differ between fields.

In an idealized form of Fano's model with several continuum
channels~\cite{Fano1961,FanoCooper1965}, a discrete state $\varphi$ is coupled
to the continuum states $\psi_{cE}$ of the channels $c=1,\dots,n_c$, which
extend over the whole energy axis. The coupling is through real,
energy-independent matrix elements, the
components of $\vec{V}_E$, without direct scattering, and the matrix elements of
$T$ do not depend on $E$. Then
$\Scc=\mathbf{1}-2\pi i\,\vec{V}_E\vec{V}_E^{\Tr}/(E-E_r+i\Gamma/2)$ with
$\Gamma=2\pi\|\vec{V}_E\|^2$, and
$\Fvec(E)=\Fbg+\vec{V}_E(\langle\varphi|T|0\rangle-i\pi\vec{V}_E^{\Tr}%
\Fbg)/(E-E_r+i\Gamma/2)$, where the components
of $\Fbg$ are the direct amplitudes $\langle\psi_{cE}|T|0\rangle$. Hence
$\sg=\Gamma_g=\Gamma$,
$\om=\pm e^{-i\pi/4}\vec{V}_E/\|\vec{V}_E\|$, and Eqs.~(\ref{eq:qg}) and
(\ref{eq:pg}) give
$\qg=\langle\varphi|T|0\rangle/(\pi\vec{V}_E^{\Tr}\Fbg)$ and
$\pg=|\vec{V}_E^{\Tr}\Fbg|/(\|\vec{V}_E\|\,\|\Fbg\|)$,
Fano's $q$ and $|\rho|$ for several continuum channels~\cite{FanoCooper1965}. In
Fano's $q$ of
Sec.~\ref{subsec:qdef}, $\langle\Phi_E|T|0\rangle=\langle\varphi|T|0\rangle$
because the continuum states extend over the whole energy axis.

\subsection{Breit--Wigner term of the resonance}
\label{subsec:BWterm}
\begin{figure*}[tp]
  \centering
  \includegraphics[width=\textwidth]{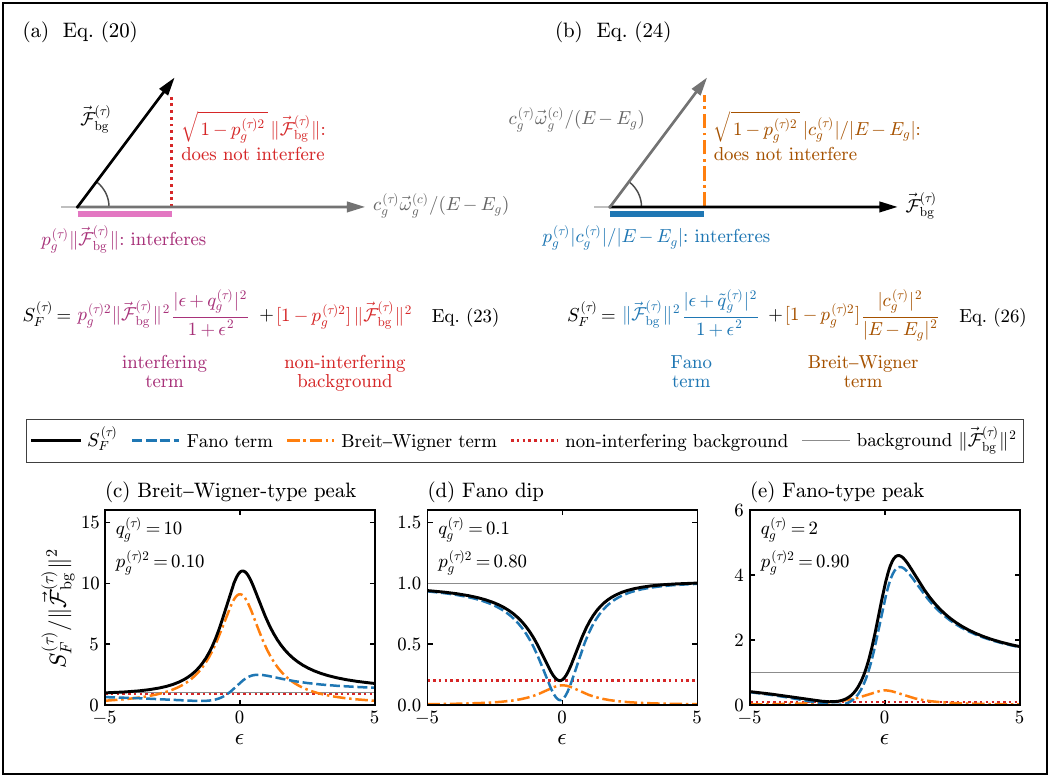}
  \caption{(a) The split~(\ref{eq:Fomega}) of the background $\Fbg$ along the
Takagi vector $\om$, on which the pole term
  $\cg\om/(E-E_g)$ lies, and (b) the split~(\ref{eq:Fpar}) of the pole term
along $\Fbg$, at one energy $E$; below each
  sketch, the resulting form of the strength, Eqs.~(\ref{eq:fano}) and
(\ref{eq:fanoBW}), with each term in the color of
  the component that gives it. Components on the same line add as amplitudes and
interfere; orthogonal components add in
  their squared norms and do not interfere. The angle is the same in (a) and
(b), with the modulus of its cosine
  $\pg$ [Eq.~(\ref{eq:pg})]; the vectors are complex, and the drawing shows
their norms and the modulus of the cosine.
  (c)--(e) The three patterns of Sec.~\ref{subsec:shapes} for constant real
$\qg$ and $p_g^{(\tau)2}$, with the
  terms of Eqs.~(\ref{eq:fanoBW}) and (\ref{eq:fano}) (legend), all divided by
$\|\Fbg\|^2$.}
  \label{fig:schematic}
\end{figure*}

Equation~(\ref{eq:Fomega}) groups the amplitude by the direction $\om$ of the
resonance. Grouped instead by the
direction of the background, it shows how much of the resonance itself
interferes. Since $\Fbgd\om=(\omd\Fbg)^*$,
Eqs.~(\ref{eq:qg}) and (\ref{eq:pg}) give the component of the pole term along
$\Fbg$ as
$p_g^{(\tau)2}(\qg-i)\Fbg/(\epsilon+i)$, so that Eq.~(\ref{eq:Fsplit}) becomes
\begin{align}
  \Fvec(E)={}&\Fbg(E)\,\frac{\epsilon+\tilde{q}_g^{(\tau)}(E)}{\epsilon+i}
  \nonumber\\
  &+\frac{1}{E-E_g}\Bigl[\mathbf{1}-\frac{\Fbg(E)\Fbgd(E)}{\|\Fbg(E)\|^2}\Bigr]%
\cg\om,
  \label{eq:Fpar}
\end{align}
with
\begin{equation}
  \tilde{q}_g^{(\tau)}(E)-i\equiv p_g^{(\tau)2}(E)\bigl[\qg(E)-i\bigr].
  \label{eq:qtilde}
\end{equation}
The first part [Fig.~\ref{fig:schematic}(b)] is the whole background times
Fano's one-continuum factor with the parameter $\tilde{q}_g^{(\tau)}$. The
second, orthogonal to it, is the rest of the pole term, of squared norm
$[1-p_g^{(\tau)2}]\,|\cg|^2/|E-E_g|^2$. The
squared norm of Eq.~(\ref{eq:Fpar}) is therefore
\begin{align}
  S_F^{(\tau)}(E)={}&\|\Fbg(E)\|^2\,\frac{|\epsilon+\tilde{q}_g^{(\tau)}(E)|^2}%
{1+\epsilon^2}
  \nonumber\\
  &+\bigl[1-p_g^{(\tau)2}(E)\bigr]\,\frac{\sg\,|\Fg|^2}{2\pi\bigl[(E-\ReEg)^2+%
\Gamma_g^2/4\bigr]},
  \label{eq:fanoBW}
\end{align}
with $|\cg|^2=\sg|\Fg|^2/2\pi$ [Eq.~(\ref{eq:resF})]. The first term is called
the Fano term: Fano's one-continuum profile
on the whole background, which contains the interference of the resonance with
the background. The second, the
Breit--Wigner term, is the part of the resonance that does not interfere. It is
a Lorentzian of width $\Gamma_g$ centered at
$\ReEg$ multiplied by $1-p_g^{(\tau)2}(E)$, the fraction of the squared norm of
the pole term orthogonal to the
background. It reduces to a Lorentzian where $\pg$ is constant across the width.

The sum of the two terms of Eq.~(\ref{eq:fanoBW}) is a single Fano profile on
the whole background with a complex
parameter whose real part is that of $\tilde{q}_g^{(\tau)}$ and whose squared
modulus is
$1-p_g^{(\tau)2}+p_g^{(\tau)2}|\qg|^2$, the form obtained for isolated
resonances in models with several open
channels~\cite{Gorin2005,Alhassid2006}. For an isolated resonance between two
single-mode leads with time-reversal symmetry, the Fano parameter is real
without decoherence~\cite{Clerk2001,Mendoza2008}, and dephasing that reinjects
the lost flux gives the same relations, with the ratio of the resonance widths
without and with dephasing in place of
$p_g^{(\tau)2}$~\cite{Clerk2001,Baernthaler2010}; dissipation alone multiplies
$q-i$ by this ratio, as $p_g^{(\tau)2}$ does in
Eq.~(\ref{eq:qtilde})~\cite{Baernthaler2010}. In Eq.~(\ref{eq:fanoBW}) the
Breit--Wigner term requires no decoherence: the strength
sums the open channels, which are orthogonal final states, without interference,
so that the part of the pole term
orthogonal to the background does not interfere with it. Channel by channel, the
amplitude of Eq.~(\ref{eq:Fomega}) has, up to complex conjugation, the form that
Ref.~\cite{Starace1977} obtained for one of several channels, with $\qg$ in
place of Starace's real $q$; the weighted mean and variance of his parameter,
the fraction of the background amplitude of a channel that passes through the
combination of continuum channels that interacts with the resonance, give the
Fano and Breit--Wigner terms.

Equations~(\ref{eq:fano}) and (\ref{eq:fanoBW}) are two exact forms of the same
strength, and they contain the same
$\pg$, the modulus of the cosine of the angle between $\Fbg$ and $\om$
[Fig.~\ref{fig:schematic}(a),(b)].
Equation~(\ref{eq:fano}) projects the background on the direction of the
resonance and separates the background that does
not interfere, and Eq.~(\ref{eq:fanoBW}) projects the resonance on the direction
of the background and separates the
part of the resonance that does not interfere. Their parameters are related by
Eq.~(\ref{eq:qtilde}). For $\pg=1$ the
two forms coincide. For $\pg\to0$, $\tilde{q}_g^{(\tau)}\to i$, the Fano term
reduces to the background, and the
resonance appears only in the Breit--Wigner term, whereas in Eq.~(\ref{eq:fano})
$|\qg|$ diverges. Equation~(\ref{eq:fano}) is the natural form for dips, since
its non-interfering background is the lower bound of the strength and $\qg$
fixes
where the strength falls below the background (Sec.~\ref{subsec:shapes}).
Equation~(\ref{eq:fanoBW}) is the natural form
for peaks, since its Breit--Wigner term shows which part of a peak does not
interfere.

\subsection{Line shapes}
\label{subsec:shapes}
Equation~(\ref{eq:fano}), rewritten as the excess over the background, is
\begin{equation}
  \frac{S_F^{(\tau)}(E)}{\|\Fbg(E)\|^2}-1=p_g^{(\tau)2}\,
  \frac{|\qg|^2-1+2\epsilon\operatorname{Re}\qg}{1+\epsilon^2}.
  \label{eq:SF-q}
\end{equation}
The non-interfering background cancels, and the strength differs from the
background only through the interfering
term. The parameter $\qg$ decides on which side of the background the strength
lies, and $p_g^{(\tau)2}$ how far. The pole
$g$ is said to produce a dip in the strength function of the field $\tau$ when
$S_F^{(\tau)}(E)$ falls below
$\|\Fbg(E)\|^2$ somewhere within a few widths of $\ReEg$ ($|\epsilon|\le4$ in
Tables~\ref{tab:cases} and
\ref{tab:casesPb}). It is said to produce a peak when $S_F^{(\tau)}/\|\Fbg\|^2$
has a maximum above one there. The comparison is made
with the background of the pole itself; it is not the same as a maximum or a
minimum of $S_F^{(\tau)}(E)$. For a real
constant $q$, Fano's profile $(\epsilon+q)^2/(1+\epsilon^2)$ has its maximum
$1+q^2$ at $\epsilon=1/q$ and its zero at
$\epsilon=-q$~\cite{Fano1961,Miroshnichenko2010}. For a constant $\qg$,
Eq.~(\ref{eq:SF-q}) is negative on the side of the zero, for
$\epsilon<(1-|\qg|^2)/(2\operatorname{Re}\qg)$ if $\operatorname{Re}\qg>0$ and
for $\epsilon$ above this value if
$\operatorname{Re}\qg<0$. When $\qg$ changes within the width, the sign of
Eq.~(\ref{eq:SF-q}) still follows pointwise
from $\qg(E)$.

The non-interfering background contains no pole term, so that
\begin{equation}
  S_F^{(\tau)}(E)\ge\bigl[1-p_g^{(\tau)2}(E)\bigr]\|\Fbg(E)\|^2,
  \label{eq:lower}
\end{equation}
with the equality where $\epsilon+\qg=0$. The resonance can cancel only the
component of the background along $\om$.
The depth of a dip relative to the background is at most $p_g^{(\tau)2}$.

Equations~(\ref{eq:fanoBW}) and (\ref{eq:SF-q}) thus give three patterns of the
strength near a pole, named as
follows and used in Secs.~\ref{sec:O16}--\ref{sec:common}
[Fig.~\ref{fig:schematic}(c)--(e)].

\emph{Breit--Wigner-type peak} [Fig.~\ref{fig:schematic}(c)]: a peak carried by
the Breit--Wigner term of
Eq.~(\ref{eq:fanoBW}), a Lorentzian added to a background with which the
resonance hardly interferes. It appears for
$p_g^{(\tau)2}\simeq0$, where $|\qg|\gg1$ with $p_g^{(\tau)2}|\qg|^2$ finite
[Eqs.~(\ref{eq:qg}) and (\ref{eq:pg})].

\emph{Fano dip} [Fig.~\ref{fig:schematic}(d)]: a dip without a peak. Since the
Breit--Wigner term is positive, a dip
comes only from the Fano term, that is, from interference. For $|\qg|<1$ the
strength at the center lies below the
background, Fano's condition for a dip at the resonance energy, and $\qg=0$
gives a symmetric dip.

\emph{Fano-type peak} [Fig.~\ref{fig:schematic}(e)]: a peak carried by the Fano
term, for $p_g^{(\tau)2}\simeq1$. It is Fano's
asymmetric profile with $|\qg|\gtrsim1$, whose zero lowers the strength on one
side to nearly $[1-p_g^{(\tau)2}]\|\Fbg\|^2$
[Eq.~(\ref{eq:lower})]. The larger $|\qg|$, the closer the peak is to a
Lorentzian and the farther the zero. For
$|\qg|\simeq1$ the maximum and the zero lie about one half-width on opposite
sides of the center, and the profile is
both a peak and a dip.

A peak at whose maximum $p_g^{(\tau)2}$ is neither close to zero nor close to
one is called mixed. Where the component of
the pole term along the background is much larger than the background at the
center, $p_g^{(\tau)2}|\qg-i|\gg1$, that
is, $|\tilde{q}_g^{(\tau)}-i|\gg1$ [Eq.~(\ref{eq:qtilde})], the Fano and
Breit--Wigner terms carry nearly the
fractions $p_g^{(\tau)2}$ and $1-p_g^{(\tau)2}$ of the strength at the center,
so that both terms contribute comparably to
a mixed peak. Otherwise the fractions can differ from these values.

\subsection{Background distribution and branching ratios}
\label{subsec:bc}
Two distributions over the open channels enter $\pg$. Since the pole term is
proportional to $\om$, the branching
ratio $\mathrm{br}_c=|\omega^{(c)}_{g,c}|^2$ of Eq.~(\ref{eq:br}) is also the
fraction of the squared norm of the pole
term carried by channel $c$. The background has the channel distribution
\begin{equation}
  b_c^{(\tau)}(E)\equiv\frac{|\mathcal{F}^{(\tau)}_{\mathrm{bg},c}(E)|^2}{\|%
\Fbg(E)\|^2},\qquad
  \sum_c b_c^{(\tau)}(E)=1,
  \label{eq:bc}
\end{equation}
which depends on $E$, on the field, and on the pole $g$ whose term has been
removed. Writing Eq.~(\ref{eq:pg}) in
components and using the triangle inequality,
\begin{equation}
  \pg(E)=\frac{\bigl|\sum_c\omega^{(c)*}_{g,c}\,\mathcal{F}^{(\tau)}_{%
\mathrm{bg},c}(E)\bigr|}{\|\Fbg(E)\|}
  \le\sum_c\sqrt{b_c^{(\tau)}(E)\,\mathrm{br}_c},
  \label{eq:pbound}
\end{equation}
with the equality when the products
$\omega^{(c)*}_{g,c}\mathcal{F}^{(\tau)}_{\mathrm{bg},c}$ of all channels have
the
same phase. The bound is the overlap of the two distributions. It equals one
only when they coincide, and it is small when the branching ratios are large
mainly in channels in which the background is small. Because $\mathrm{br}_c$
does not depend on the field
while $b_c^{(\tau)}$ does, the same pole can interfere with the backgrounds of
different fields to different degrees.

\subsection{Phase of the transition amplitude and the collectivity indicators}
\label{subsec:phase}
The parameter $\qg$ of Eq.~(\ref{eq:qg}) is defined on the real axis, whereas
the complex
transition probability~(\ref{eq:Theta}) is a residue at the pole. They are
related through the background amplitude
along $\om$, $\omd\Fbg(E)$, which is analytic in the upper half plane of the
physical sheet. Equation~(\ref{eq:qg})
with $\omd\Fbg$ taken at $E_g^*=\ReEg+i\Gamma_g/2$, half a width above the real
axis, defines $\bar{q}_g^{(\tau)}$, which is called Fano's parameter at the
pole. Comparing the residues at $E_g$ of the two forms of the strength function,
$-\operatorname{Im}R_F^{(\tau)}/\pi$
and Eq.~(\ref{eq:SF-norm}), both continued to complex energies, gives
(Appendix~\ref{app:phase})
\begin{equation}
  \widetilde{B}_g^{(\tau)}=\frac{\sg}{\Gamma_g}\,|\Fg|^2\,
  \frac{\bar{q}_g^{(\tau)*}-i}{\bar{q}_g^{(\tau)*}+i},
  \label{eq:Bres}
\end{equation}
and, with Eq.~(\ref{eq:Theta}),
\begin{equation}
  \bar{q}_g^{(\tau)}=i\,\frac{\Gamma_g+\sg e^{2i\Theta_g^{(\tau)}}}{\Gamma_g-\sg
e^{2i\Theta_g^{(\tau)}}}.
  \label{eq:qTheta}
\end{equation}
At the pole, Fano's parameter is thus fixed by the phase of the transition
amplitude and by $\sg/\Gamma_g$. For
$\sg=\Gamma_g$ it is real, $\bar{q}_g^{(\tau)}=-\cot\Theta_g^{(\tau)}$, the
parameter that Ref.~\cite{Mizuyama2026_Collectivity}
obtained by matching Fano's profile to the pole term of the strength function.
The relation $\bar{q}_g^{(\tau)}=-\cot\Theta_g^{(\tau)}$ is the one found in
Ref.~\cite{Ott2013} between Fano's parameter and the phase of the dipole
response, with that phase equal to $2\Theta_g^{(\tau)}$, the phase of
$\widetilde{B}_g^{(\tau)}$. In particle physics, the shape of a resonance peak
has likewise been described by the phase of the pole residue relative to the
background~\cite{Ceci2013}.

Writing Eq.~(\ref{eq:Bres}) as
$\widetilde{B}_g^{(\tau)}=(\sg/\Gamma_g)|\Fg|^2(|\bar{q}_g^{(\tau)}|^2-1-2i%
\operatorname{Re}\bar{q}_g^{(\tau)})/|\bar{q}_g^{(\tau)}-i|^2$
shows that $\operatorname{Re}\widetilde{B}_g^{(\tau)}<0$ is equivalent to
$|\bar{q}_g^{(\tau)}|<1$, that is, to $\pi/4<|\Theta_g^{(\tau)}|<3\pi/4$. In
Refs.~\cite{Mizuyama2025_JRPA2,Mizuyama2026_Collectivity} a negative real part
of the complex transition probability was
associated with a possible Fano dip; it is Fano's condition $|q|<1$ for a dip at
the resonance energy
(Sec.~\ref{subsec:shapes}), evaluated at $E_g^*$. With $\qg$ and $\omd\Fbg$
fixed at their values at $E_g^*$,
the right side of Eq.~(\ref{eq:SF-q}) times $\|\Fbg\|^2$ becomes the
contribution of the pole term of the response
function, $-\operatorname{Im}[\widetilde{B}_g^{(\tau)}/(E-E_g)]/\pi$, the
single-pole term of
Refs.~\cite{Mizuyama2025_JRPA2,Mizuyama2026_Collectivity}. This term consists of
a Lorentzian and a dispersive part, Shore's
parameterization, equivalent to Fano's profile~\cite{Shore1967}.

The phase does not fix $\pg$, which sets the pattern of a peak and bounds the
depth of a dip [Eq.~(\ref{eq:lower})], since $\pg$ depends on the direction of
the background relative to $\om$. Moreover, $\qg(\ReEg)$ differs from
$\bar{q}_g^{(\tau)}$
when the background changes over the distance $\Gamma_g/2$ between the real axis
and $E_g^*$; Tables~\ref{tab:cases}
and \ref{tab:casesPb} give both.

The collectivity indicators enter the strength only through the size of the pole
term. By Eqs.~(\ref{eq:resF}) and (\ref{eq:Bfact}), the squared norm of the
residue is
\begin{equation}
  |\cg|^2=\frac{\sg}{2\pi}\,|\Fg|^2
  =\frac{\sg}{2\pi}\,C_g^{(\tau)2}\Ntot\eta_g^{(\tau)}\sum_{i=1}^{\Ntot}|%
\Fgi|^2,
  \label{eq:c2coll}
\end{equation}
so that the channel amplitudes enter the line shape only through $|\Fg|$ and the
phase $\Theta_g^{(\tau)}$. The indicators fix neither $\pg$ nor $\qg$, nor
$\bar{q}_g^{(\tau)}$, although they enter the modulus
$|\widetilde{B}_g^{(\tau)}|=|\Fg|^2$. By Eq.~(\ref{eq:qTheta}),
$|\bar{q}_g^{(\tau)}-i|=2/|\Gamma_g/\sg-e^{2i\Theta_g^{(\tau)}}|$: the ratio of
the pole term at the center, $|\cg|/(\Gamma_g/2)$, to the background along $\om$
at $E_g^*$ [Eq.~(\ref{eq:qg})] does not depend on $|\Fg|$.

\section{Poles of $^{16}$O}
\label{sec:O16}
The model, its parameters, and the energy grids are given in
Appendix~\ref{app:jrpa}. For each pole, $\Fbg(E)$ is
obtained from $\Fvec(E)$ on the real interval of its sheet, or on a part of it,
by Eq.~(\ref{eq:Fsplit}), and the
residue is checked by a contour integral. The energy range of the panels that
show the decomposition for a pole is
called its window; the whole-strength panels of each figure mark it.
Tables~\ref{tab:cases} and \ref{tab:casesPb} list
the cases analyzed in this section and in Sec.~\ref{sec:Pb}, except the poles
whose terms are only removed in Secs.~\ref{subsec:dipO16} and
\ref{subsec:dipPb}, and give under each case the figure that shows
it. The cases are examples of the three patterns of Sec.~\ref{subsec:shapes} in
a light and a heavy nucleus, including
giant resonances and narrow poles beside them; they are not a survey of all the
poles found.

\squeezetable
\begin{table*}[tbp]
\caption{Poles of $^{16}$O analyzed with Fano's formula~(\ref{eq:fano}) for the
isoscalar (IS, $\tau=0$) and isovector (IV, $\tau=1$) fields. $E_g$ is in MeV,
and $\operatorname{Re}\widetilde{B}_g^{(\tau)}$ in fm$^4$ for $2^+$, fm$^6$ for
$1^-$ IS, and fm$^2$ for $1^-$ IV. The numbers under the column heads are those
of the equations that define the quantities; $\qg$ and $p_g^{(\tau)2}$ are given
at the grid point nearest $\ReEg$. Result: the patterns of
Sec.~\ref{subsec:shapes} [Fig.~\ref{fig:schematic}(c)--(e)], judged within the
computed part of $|\epsilon|\le4$: a Fano dip if $S_F^{(\tau)}<\|\Fbg\|^2$ there
without a maximum of $S_F^{(\tau)}/\|\Fbg\|^2$ above one; a peak is a Fano-type,
a Breit--Wigner-type (BW-type), or a mixed peak for $p_g^{(\tau)2}$ at the
maximum of $S_F^{(\tau)}/\|\Fbg\|^2$ close to one, close to zero, or in between;
``and dip'' if the strength also falls below the background. Under each case,
the figure that shows it.}
\label{tab:cases}
\begin{ruledtabular}
\begin{tabular}{llccccccc}
Case & \begin{tabular}{@{}c@{}}$E_g$\\(\ref{eq:takagi})\end{tabular} &
\begin{tabular}{@{}c@{}}$C_g^{(\tau)2}$\\$\times\Ntot\eta_g^{(\tau)}$\\(%
\ref{eq:CNeta})\end{tabular} &
\begin{tabular}{@{}c@{}}$\Theta_g^{(\tau)}/\pi$\\(\ref{eq:Theta})\end{tabular} &
\begin{tabular}{@{}c@{}}$\operatorname{Re}\widetilde{B}_g^{(\tau)}$\\(%
\ref{eq:Btilde})\end{tabular} &
\begin{tabular}{@{}c@{}}$\bar{q}_g^{(\tau)}$\\(\ref{eq:qTheta})\end{tabular} &
\begin{tabular}{@{}c@{}}$\qg$\\(\ref{eq:qg})\end{tabular} &
\begin{tabular}{@{}c@{}}$p_g^{(\tau)2}$\\(\ref{eq:pg})\end{tabular} & Result \\
\hline
\begin{tabular}{@{}c@{}}$2^+$, IS\\Fig.~\ref{fig:dipO16}\end{tabular} &
$28.495-0.564i$ & 0.0232 & $0.617$ & $-0.0879$ & $0.39-0.05i$ & $0.34+0.00i$ &
0.234 & \begin{tabular}{@{}c@{}}Fano\\dip\end{tabular} \\
\hline
\begin{tabular}{@{}c@{}}$1^-$, IS\\Fig.~\ref{fig:peak15794}, left\end{tabular} &
$15.794-0.003i$ & 0.940 & $0.007$ & $4.84$ & $-43.5-4.1i$ & $-43.7-3.6i$ & 0.110
& \begin{tabular}{@{}c@{}}BW-type\\peak\end{tabular} \\
\begin{tabular}{@{}c@{}}$1^-$, IV\\Fig.~\ref{fig:peak15794}, right\end{tabular}
&  & 0.0175 & $-0.984$ & $0.00637$ & $-19.4-0.8i$ & $-19.4-0.8i$ & 0.895 &
\begin{tabular}{@{}c@{}}Fano-type\\peak\end{tabular} \\
\hline
\begin{tabular}{@{}c@{}}$2^+$, IS\\Fig.~\ref{fig:peak16762}, left\end{tabular} &
$16.762-0.107i$\footnotemark[1] & 7.09 & $0.993$ & $82.9$ & $26.1-23.9i$ &
$24.1-24.5i$ & 0.096 & \begin{tabular}{@{}c@{}}BW-type\\peak\end{tabular} \\
\begin{tabular}{@{}c@{}}$2^+$, IV\\Fig.~\ref{fig:peak16762}, right\end{tabular}
&  & 0.00769 & $0.306$ & $-0.0312$ & $-0.70-0.03i$ & $-0.74+0.03i$ & 0.526 &
\begin{tabular}{@{}c@{}}mixed\\peak\\and dip\end{tabular} \\
\end{tabular}
\end{ruledtabular}
\footnotetext[1]{Pole $16.762-0.107i$~MeV: the computed interval,
16.43--16.89~MeV, is the whole interval of the sheet and covers
$-3.05\le\epsilon\le1.16$.}
\end{table*}
\squeezetable
\begin{table*}[tbp]
\caption{As Table~\ref{tab:cases}, for the $1^-$ poles of $^{208}$Pb; for the
poles $21.142-0.373i$ and $21.275-0.020i$~MeV only the isoscalar field is
given.}
\label{tab:casesPb}
\begin{ruledtabular}
\begin{tabular}{llccccccc}
Field & \begin{tabular}{@{}c@{}}$E_g$\\(\ref{eq:takagi})\end{tabular} &
\begin{tabular}{@{}c@{}}$C_g^{(\tau)2}$\\$\times\Ntot\eta_g^{(\tau)}$\\(%
\ref{eq:CNeta})\end{tabular} &
\begin{tabular}{@{}c@{}}$\Theta_g^{(\tau)}/\pi$\\(\ref{eq:Theta})\end{tabular} &
\begin{tabular}{@{}c@{}}$\operatorname{Re}\widetilde{B}_g^{(\tau)}$\\(%
\ref{eq:Btilde})\end{tabular} &
\begin{tabular}{@{}c@{}}$\bar{q}_g^{(\tau)}$\\(\ref{eq:qTheta})\end{tabular} &
\begin{tabular}{@{}c@{}}$\qg$\\(\ref{eq:qg})\end{tabular} &
\begin{tabular}{@{}c@{}}$p_g^{(\tau)2}$\\(\ref{eq:pg})\end{tabular} & Result \\
\hline
\begin{tabular}{@{}c@{}}IS\\Fig.~\ref{fig:pairPbIS}, left\end{tabular} &
$21.142-0.373i$\footnotemark[1] & 11.9 & $0.135$ & $3.50\times10^{4}$ &
$-1.57-1.14i$ & $-0.62-1.02i$ & 0.891 &
\begin{tabular}{@{}c@{}}Fano-type\\peak\\and dip\end{tabular} \\
\hline
\begin{tabular}{@{}c@{}}IS\\Fig.~\ref{fig:pairPbIS}, right\end{tabular} &
$21.275-0.020i$ & 0.285 & $-0.233$ & $224$ & $1.11-0.12i$ & $0.97-0.10i$ & 0.813
& \begin{tabular}{@{}c@{}}Fano-type\\peak\\and dip\end{tabular} \\
\hline
\begin{tabular}{@{}c@{}}IS\\Fig.~\ref{fig:pairPbIS12}, left\end{tabular} &
$13.181-0.105i$ & 0.0852 & $-0.493$ & $-146$ & $0.02-0.00i$ & $0.12+0.01i$ &
0.388 & \begin{tabular}{@{}c@{}}Fano\\dip\end{tabular} \\
\begin{tabular}{@{}c@{}}IV\\Fig.~\ref{fig:pairPbIV}, left\end{tabular} &  & 6.00
& $0.969$ & $2.87$ & $10.3-0.1i$ & $8.83-2.35i$ & 0.741 &
\begin{tabular}{@{}c@{}}Fano-type\\peak\\and dip\end{tabular} \\
\hline
\begin{tabular}{@{}c@{}}IS\\Fig.~\ref{fig:pairPbIS12}, right\end{tabular} &
$13.412-0.007i$ & 3.22 & $0.066$ & $323$ & $-4.71-0.24i$ & $-4.66-0.25i$ & 0.655
& \begin{tabular}{@{}c@{}}mixed\\peak\\and dip\end{tabular} \\
\begin{tabular}{@{}c@{}}IV\\Fig.~\ref{fig:pairPbIV}, right\end{tabular} &  &
3.29 & $-0.117$ & $0.278$ & $2.59-0.08i$ & $2.53-0.12i$ & 0.940 &
\begin{tabular}{@{}c@{}}Fano-type\\peak\\and dip\end{tabular} \\
\end{tabular}
\end{ruledtabular}
\footnotetext[1]{Pole $21.142-0.373i$~MeV: the decomposition covers the interval
of $\mathrm{RS}=23$, 20.63--21.95~MeV ($-1.38\le\epsilon\le2.17$).}
\end{table*}

\subsection{The isoscalar $2^+$ dip near 28~MeV}
\label{subsec:dipO16}
\begin{figure*}[tp]
  \centering
  \includegraphics[width=\textwidth]{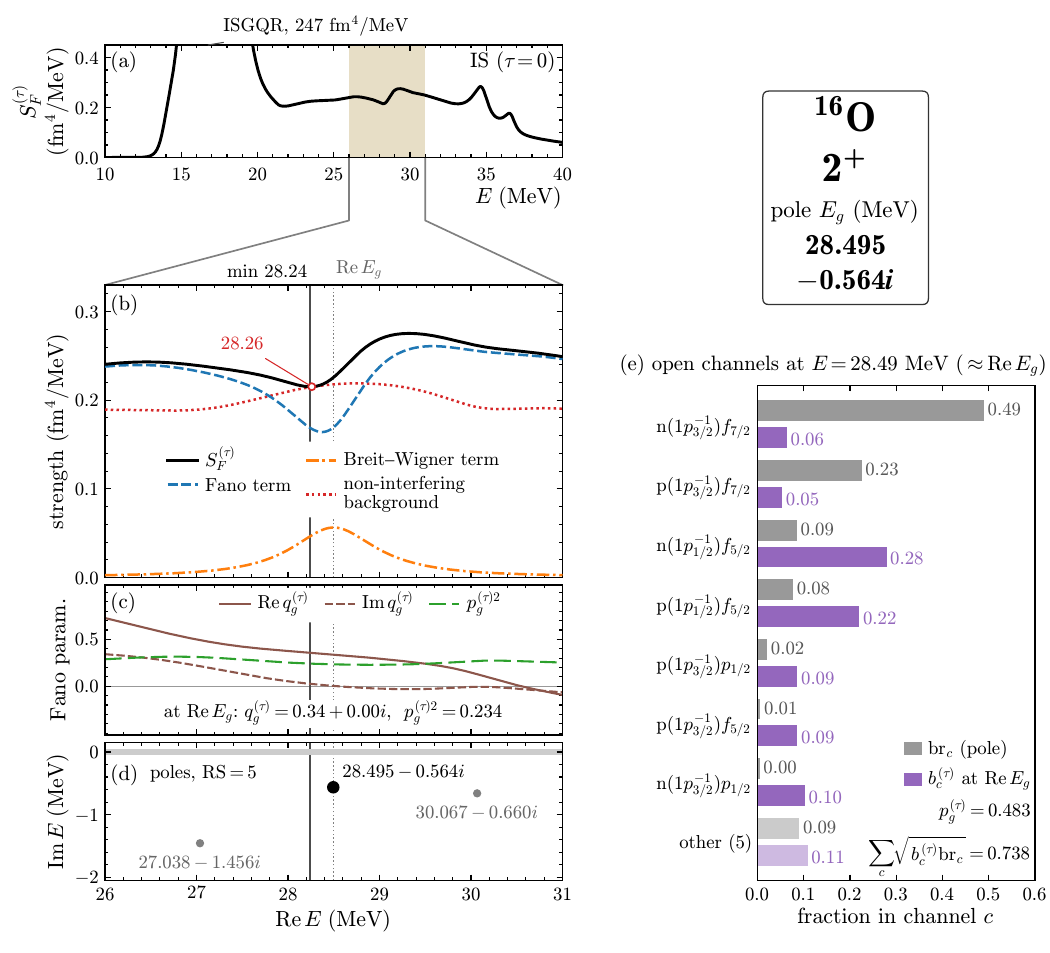}
  \caption{Isoscalar ($\tau=0$) strength of $^{16}$O $2^+$ and the pole
$E_g=28.495-0.564i$~MeV ($\mathrm{RS}=5$) in
  the window 26--31~MeV. (a) Whole strength, with the window shaded and the
height of the off-scale ISGQR peak. (b)
  Decomposition into the terms of Eqs.~(\ref{eq:fanoBW}) and (\ref{eq:fano})
(legend); the non-interfering
  background is the lower bound~(\ref{eq:lower}); open circle: minimum of
$S_F^{(\tau)}/\|\Fbg\|^2$. (c) Fano
  parameters $\qg$ [Eq.~(\ref{eq:qg})] and $p_g^{(\tau)2}$ [Eq.~(\ref{eq:pg})].
(d) Poles of the sheet, the analyzed
  pole black and the others gray; thick line: real-energy interval of the sheet.
(e) Branching ratios
  $\mathrm{br}_c$ [Eq.~(\ref{eq:br})] of the pole and background distribution
$b_c^{(\tau)}$ [Eq.~(\ref{eq:bc})] at
  $\ReEg$, with $\pg$ and its bound~(\ref{eq:pbound}), for the channels with
$\mathrm{br}_c$ or
  $b_c^{(\tau)}\ge0.05$; ``other'': the remaining open channels. Vertical lines:
$\ReEg$ (dotted) and the minimum of
  $S_F^{(\tau)}$ below the background (solid). Values at $\ReEg$ are taken at
the grid point nearest to it.}
  \label{fig:dipO16}
\end{figure*}

The isoscalar $2^+$ strength of $^{16}$O is dominated by the ISGQR, the pole
$16.762-0.107i$~MeV, whose peak reaches
247~fm$^4$/MeV at 16.764~MeV [Fig.~\ref{fig:dipO16}(a)]. Above 21~MeV the
strength is smaller by three orders of
magnitude, and between 25 and 32~MeV it has a single minimum, at 28.24~MeV.
Three poles of $\mathrm{RS}=5$, whose
real-energy interval 21.35--31.22~MeV contains the window 26--31~MeV, lie near
it, $27.038-1.456i$, $28.495-0.564i$,
and $30.067-0.660i$~MeV [Fig.~\ref{fig:dipO16}(d)]. These poles and the ISGQR
were also obtained in
Ref.~\cite{Mizuyama2026_Collectivity}, with an earlier version of the code
($\ReEg$ within 0.01~MeV and $\Gamma_g/2$
within 0.03~MeV of the present values).

For the pole $28.495-0.564i$~MeV, $q_g^{(0)}=0.34+0.00i$ and $p_g^{(0)2}=0.234$
at $\ReEg$, and $q_g^{(0)}$ stays nearly real within the width
[Fig.~\ref{fig:dipO16}(c)]. The pole produces a Fano dip
[Fig.~\ref{fig:dipO16}(b)]: with $|q_g^{(0)}|<1$ the
strength at the center is 0.79 of the background, and it stays below the
background over the whole window
($-4.42\le\epsilon\le4.44$). The minimum of
$S_F^{(0)}/\|\vec{\mathcal{F}}^{(0)}_{\mathrm{bg}}\|^2$, 0.760 at
28.26~MeV ($\epsilon=-0.42$), lies near the zero of Fano's profile,
$\epsilon=-\operatorname{Re}q_g^{(0)}=-0.34$, and
within 0.001 of the lower bound $1-p_g^{(0)2}$ there. At this minimum the
resonance cancels nearly the whole component
of the background along $\om$, and the depth is limited by $p_g^{(0)2}$.

The minimum of $S_F^{(0)}$ itself is assigned to a pole by removing pole terms
from $\Fvec$ [Eq.~(\ref{eq:Fsplit})]. Without the
term of the pole at 28.495~MeV, the strength is nearly flat, between 0.274 and
0.285~fm$^4$/MeV over 26.5--29.5~MeV.
Without the term of the pole at 27.038 or 30.067~MeV instead, a minimum remains,
at 28.10 or 28.26~MeV. The minimum at
28.24~MeV is thus produced by the pole at 28.495~MeV, in agreement with
Ref.~\cite{Mizuyama2026_Collectivity}.

The value of $p_g^{(0)}$ is limited by the channel distributions
[Fig.~\ref{fig:dipO16}(e)]. The branching ratios of the pole are largest in the
channels $\mathrm{n}(1p_{3/2}^{-1})f_{7/2}$ and
$\mathrm{p}(1p_{3/2}^{-1})f_{7/2}$ ($\mathrm{br}_c=0.490$ and
0.225), whereas the isoscalar background at $\ReEg$ is largest in
$\mathrm{n}(1p_{1/2}^{-1})f_{5/2}$ and
$\mathrm{p}(1p_{1/2}^{-1})f_{5/2}$ ($b_c^{(0)}=0.279$ and 0.220). The bound of
Eq.~(\ref{eq:pbound}) is 0.738, and the
relative phases lower $p_g^{(0)}$ to 0.483.

For the pole at 28.50~MeV, Ref.~\cite{Mizuyama2026_Collectivity} obtained
$q=-\cot\Theta_g^{(0)}=0.385$ by matching Fano's profile to the single-pole
term. This $q$ is Fano's parameter at the pole for $\sg=\Gamma_g$
(Sec.~\ref{subsec:phase}); with the present $\sg/\Gamma_g$,
Eq.~(\ref{eq:qTheta}) gives $\bar{q}_g^{(0)}=0.39-0.05i$, close to
$q_g^{(0)}=0.34$ at $\ReEg$, and $|\bar{q}_g^{(0)}|<1$ is equivalent to the
negative real part of $\widetilde{B}_g^{(0)}$ found in that reference.

\subsection{The narrow pole at 15.794~MeV}
\label{subsec:narrow}
\begin{figure*}[tp]
  \centering
  \includegraphics[width=\textwidth]{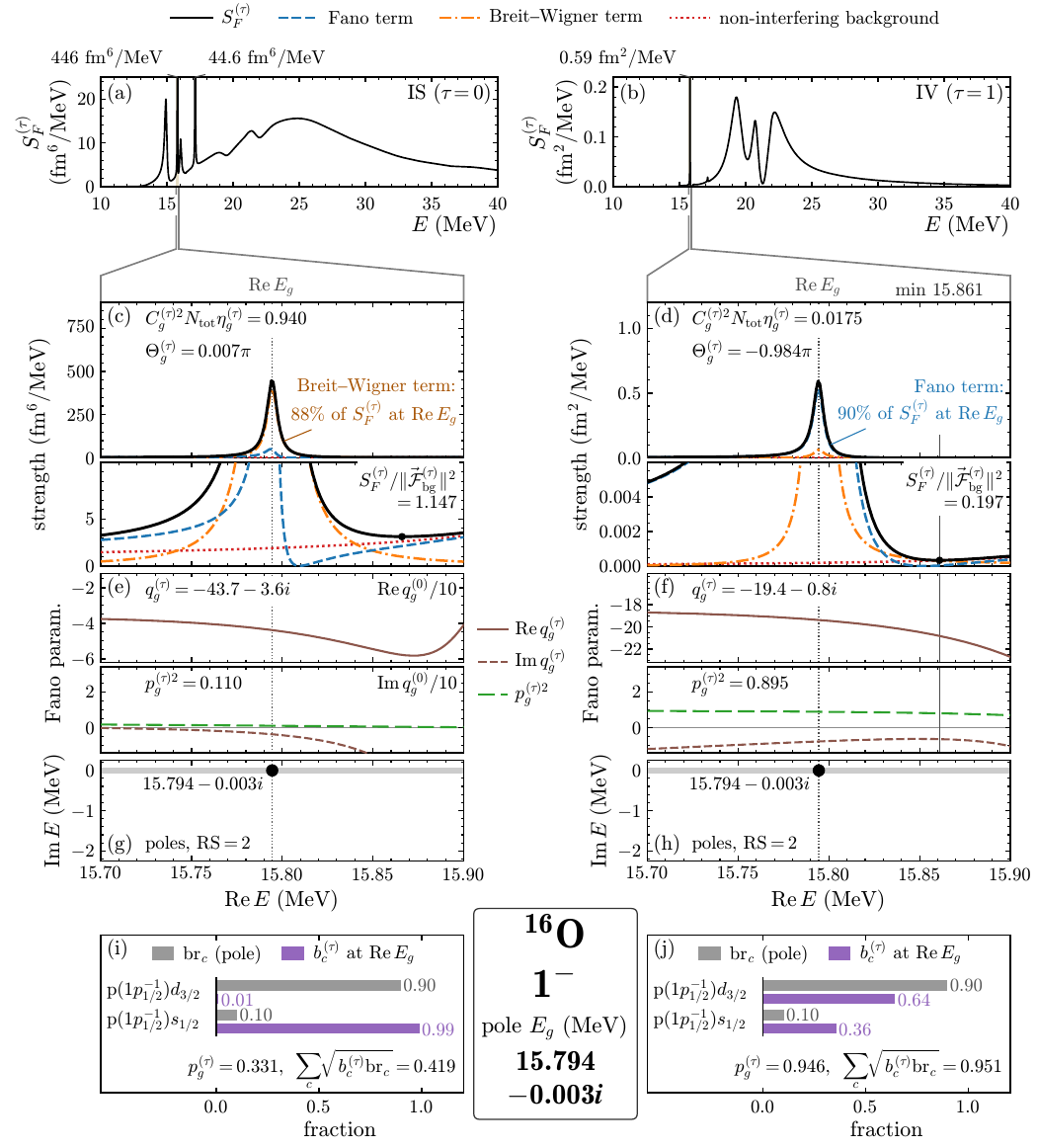}
  \caption{As Fig.~\ref{fig:dipO16}, for the pole $E_g=15.794-0.003i$~MeV of
$^{16}$O $1^-$, with the IS (left) and
  IV (right) fields; the rows (c),(d) to (i),(j) correspond to
Fig.~\ref{fig:dipO16}(b), (c), (d), and (e). (a),(b)
  Whole strength on the 10-keV grid, and on 12.00--16.43~MeV on the 1-keV grid,
which resolves the narrow peak;
  peaks cut by the frame are labeled with their heights, and the narrow band is
the window. (c),(d) Full height
  (upper part) and magnified to the level of the background (lower part; dot:
minimum of $S_F^{(\tau)}$, with
  $S_F^{(\tau)}/\|\Fbg\|^2$ there). The labels in (c),(d) give
$C_g^{(\tau)2}\Ntot\eta_g^{(\tau)}$
  [Eq.~(\ref{eq:CNeta})], the phase $\Theta_g^{(\tau)}$ [Eq.~(\ref{eq:Theta})],
and the share of the larger term of
  Eq.~(\ref{eq:fanoBW}) at $\ReEg$, and those in (e),(f) give $\qg$ and
$p_g^{(\tau)2}$ at $\ReEg$.}
  \label{fig:peak15794}
\end{figure*}

A peak in the strength function does not show by itself whether the resonance
interferes with the background. Its
pattern is set by $p_g^{(\tau)2}$ at its maximum (Sec.~\ref{subsec:shapes},
Fig.~\ref{fig:schematic}): a
Breit--Wigner-type peak for $p_g^{(\tau)2}\simeq0$, a Fano-type peak for
$p_g^{(\tau)2}\simeq1$, and a mixed peak in
between. The $1^-$ pole $15.794-0.003i$~MeV of $^{16}$O ($\Gamma_g=6.9$~keV)
lies on $\mathrm{RS}=2$ (11.998--16.434~MeV), where
only the channels $\mathrm{p}(1p_{1/2}^{-1})s_{1/2}$ and
$\mathrm{p}(1p_{1/2}^{-1})d_{3/2}$ are open. It produces the
highest peak of both the isoscalar and the isovector $1^-$ strengths
(Fig.~\ref{fig:peak15794}). In both fields
$|\qg|$ is large, $q_g^{(0)}=-43.7-3.6i$ and $q_g^{(1)}=-19.4-0.8i$ at $\ReEg$,
and both peaks are nearly Lorentzian.
Their patterns nevertheless differ. For the isoscalar field, $p_g^{(0)2}=0.110$,
and the Breit--Wigner term of
Eq.~(\ref{eq:fanoBW}) carries 88\% of the strength at $\ReEg$: a
Breit--Wigner-type peak. For the isovector field,
$p_g^{(1)2}=0.895$, and the Fano term carries 90\%: a Fano-type peak. The
difference appears on the high-energy side,
where Fano's zero lies. The isovector strength falls below the background from
15.829~MeV ($\epsilon=10$), outside
$|\epsilon|\le4$, to the end of the computed interval, and its minimum, 0.197 of
the background, is close to
$1-p_g^{(1)2}=0.192$ there. The isoscalar strength stays above the background
over the whole window 15.70--15.90~MeV.

The two values of $\pg$ are limited by the channel distributions
[Fig.~\ref{fig:peak15794}(i),(j)]. The branching ratio of the pole is largest in
$\mathrm{p}(1p_{1/2}^{-1})d_{3/2}$ ($\mathrm{br}_c=0.898$). The isoscalar
background at $\ReEg$ lies almost
entirely in the $s_{1/2}$ channel ($b_c^{(0)}=0.988$); the bound of
Eq.~(\ref{eq:pbound}) is 0.419, and
$p_g^{(0)}=0.331$. The isovector background lies mainly in the $d_{3/2}$ channel
($b_c^{(1)}=0.644$); the bound is
0.951, and $p_g^{(1)}=0.946$.

\subsection{The ISGQR}
\label{subsec:isgqr}
\begin{figure*}[tp]
  \centering
  \includegraphics[width=\textwidth]{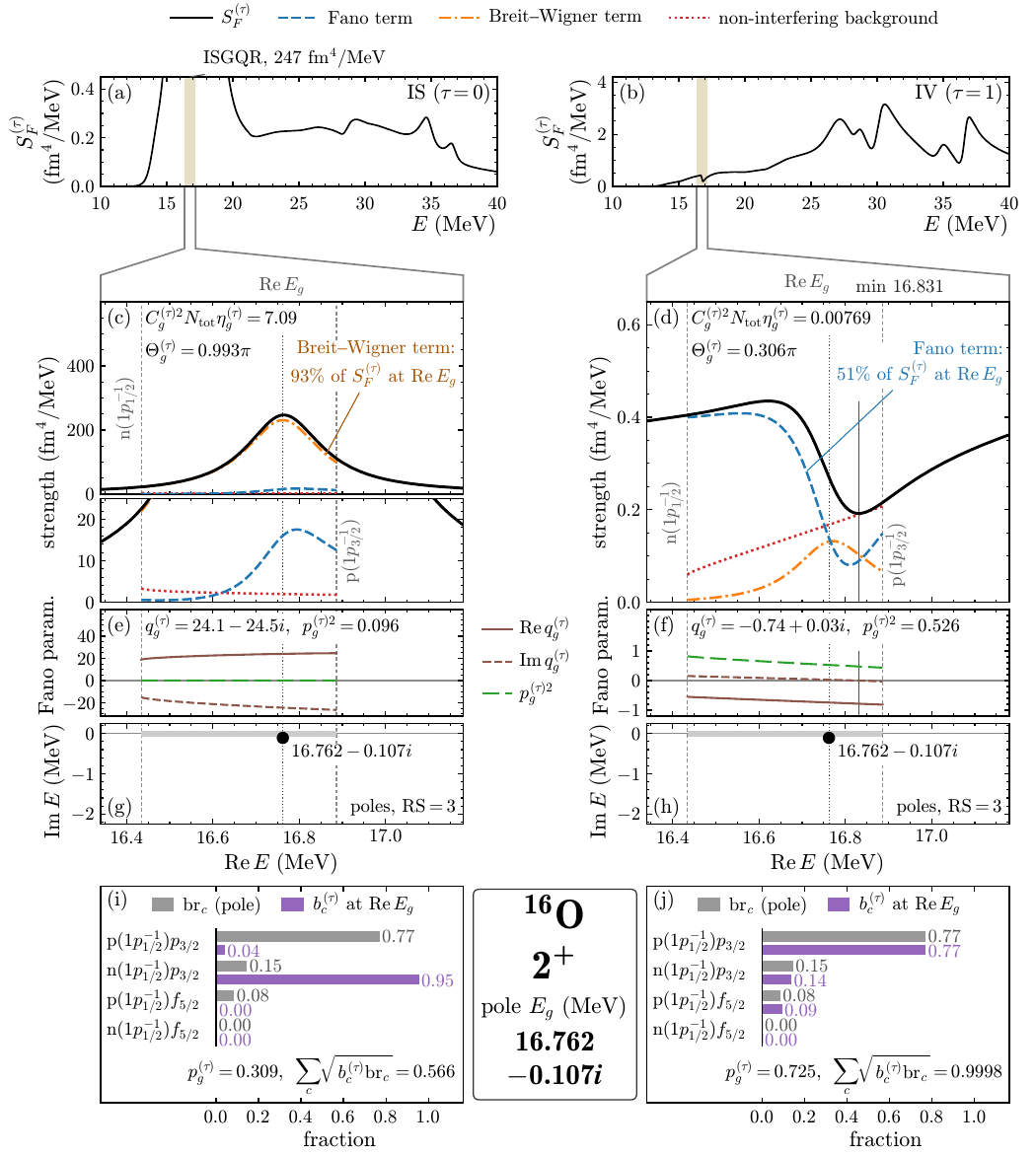}
  \caption{As Fig.~\ref{fig:peak15794}, for the ISGQR pole
$E_g=16.762-0.107i$~MeV of $^{16}$O $2^+$, with (a),(b)
  on the 10-keV grid only, (c) in two parts without a dot, (d)--(f) in one part
each, and (i),(j) for all four open
  channels. The decomposition covers the interval of $\mathrm{RS}=3$ between the
$\mathrm{n}(1p_{1/2}^{-1})$ and
  $\mathrm{p}(1p_{3/2}^{-1})$ thresholds (dashed); outside it only
$S_F^{(\tau)}$ is shown.}
  \label{fig:peak16762}
\end{figure*}

The ISGQR pole $16.762-0.107i$~MeV of $^{16}$O lies on $\mathrm{RS}=3$
(16.434--16.886~MeV, $n_c=4$), between the
$\mathrm{n}(1p_{1/2}^{-1})$ and $\mathrm{p}(1p_{3/2}^{-1})$ thresholds
(Fig.~\ref{fig:peak16762}). Its isoscalar
indicator, $C_g^{(0)2}\Ntot\eta_g^{(0)}=7.09$, is the largest among the poles of
$^{16}$O in Table~\ref{tab:cases}. In the isoscalar field the pole produces a
Breit--Wigner-type peak: at $\ReEg$, $p_g^{(0)2}=0.096$
and $|q_g^{(0)}|=34$, and the Breit--Wigner term carries 93\% of the strength.
The branching ratio of the pole is largest in $\mathrm{p}(1p_{1/2}^{-1})p_{3/2}$
($\mathrm{br}_c=0.770$), whereas the isoscalar background lies almost entirely
in
$\mathrm{n}(1p_{1/2}^{-1})p_{3/2}$ ($b_c^{(0)}=0.955$); the bound is 0.566, and
$p_g^{(0)}=0.309$.

In the isovector field the same pole produces the dip near the peak energy found
in
Ref.~\cite{Mizuyama2025_JRPA2}. At $\ReEg$, $q_g^{(1)}=-0.74+0.03i$ and
$p_g^{(1)2}=0.526$. The strength lies below the background from 16.727~MeV
($\epsilon=-0.32$) to the upper edge of the sheet, a region that contains the
minimum of $S_F^{(1)}$ itself, at 16.831~MeV. The minimum of
$S_F^{(1)}/\|\vec{\mathcal{F}}^{(1)}_{\mathrm{bg}}\|^2$, 0.531 at 16.834~MeV, is
close to $1-p_g^{(1)2}=0.527$ there.
On the low-energy side the ratio reaches 1.28 at 16.606~MeV, where
$p_g^{(1)2}=0.65$: a weak mixed peak beside the dip. The ISGQR is the only pole
found on $\mathrm{RS}=3$, and without its term the strength,
$\|\vec{\mathcal{F}}^{(1)}_{\mathrm{bg}}\|^2$, increases monotonically over the
sheet, so that the minimum of $S_F^{(1)}$ is produced by this pole.
The isovector background is distributed over the channels as the branching
ratios are ($b_c^{(1)}=0.769$, 0.137, and
0.093 in the three channels with the largest $\mathrm{br}_c$); the bound is
0.9998, and the relative phases lower
$p_g^{(1)}$ to 0.725. At the pole, Eq.~(\ref{eq:qTheta}) gives
$\bar{q}_g^{(1)}=-0.70-0.03i$, so that
$\operatorname{Re}\widetilde{B}_g^{(1)}<0$.

\section{Poles of $^{208}$Pb}
\label{sec:Pb}
The isoscalar $1^-$ strength of $^{208}$Pb has the isoscalar giant dipole
resonance (ISGDR), whose strength in the
continuum~\cite{GorelikUrin2001,HamamotoSagawa2002} and direct nucleon
decay~\cite{GorelikUrin2001} have been studied with continuum approaches. In
this model the ISGDR is the pole $21.142-0.373i$~MeV on $\mathrm{RS}=23$
(20.626--21.954~MeV, $n_c=62$) [Fig.~\ref{fig:pairPbIS}(a)], which has the
largest isoscalar indicator of all the poles found. The isovector strength has
the isovector giant
dipole resonance (IVGDR). In this model it appears as three poles of
$\mathrm{RS}=12$ (11.184--14.607~MeV, $n_c=31$),
$13.181-0.105i$, $13.412-0.007i$, and $13.697-0.019i$~MeV, whose coherence
indices $C_g^{(1)2}=0.44$, 0.29, and 0.59
separate them from all the other poles found ($C_g^{(1)2}\le0.16$); the narrow
peaks between 9 and 12.5~MeV, which
carry most of the isovector strength, come from poles of low coherence. The
broadest of the three, $13.181-0.105i$~MeV,
is called the IVGDR pole; its peak reaches 8.37~fm$^2$/MeV at 13.19~MeV
[Fig.~\ref{fig:pairPbIV}(a)]. Above each of the ISGDR and IVGDR poles lies a
narrow pole of the same sheet, $21.275-0.020i$~MeV ($\Gamma_g=40$~keV) and
$13.412-0.007i$~MeV ($\Gamma_g=15$~keV), next to which the strength
has a narrow minimum followed by a narrow peak [Figs.~\ref{fig:pairPbIS}(d) and
\ref{fig:pairPbIV}(d)]. The interplay
between a giant resonance and the narrow levels underlying it was studied in the
continuum shell model in
Ref.~\cite{Kleinwachter1986}.
The ISGDR and IVGDR poles are discussed in Sec.~\ref{subsec:giant} and the
narrow poles in Sec.~\ref{subsec:dipPb}.

\subsection{The ISGDR and IVGDR poles}
\label{subsec:giant}
\begin{figure*}[tp]
  \centering
  \includegraphics[width=\textwidth]{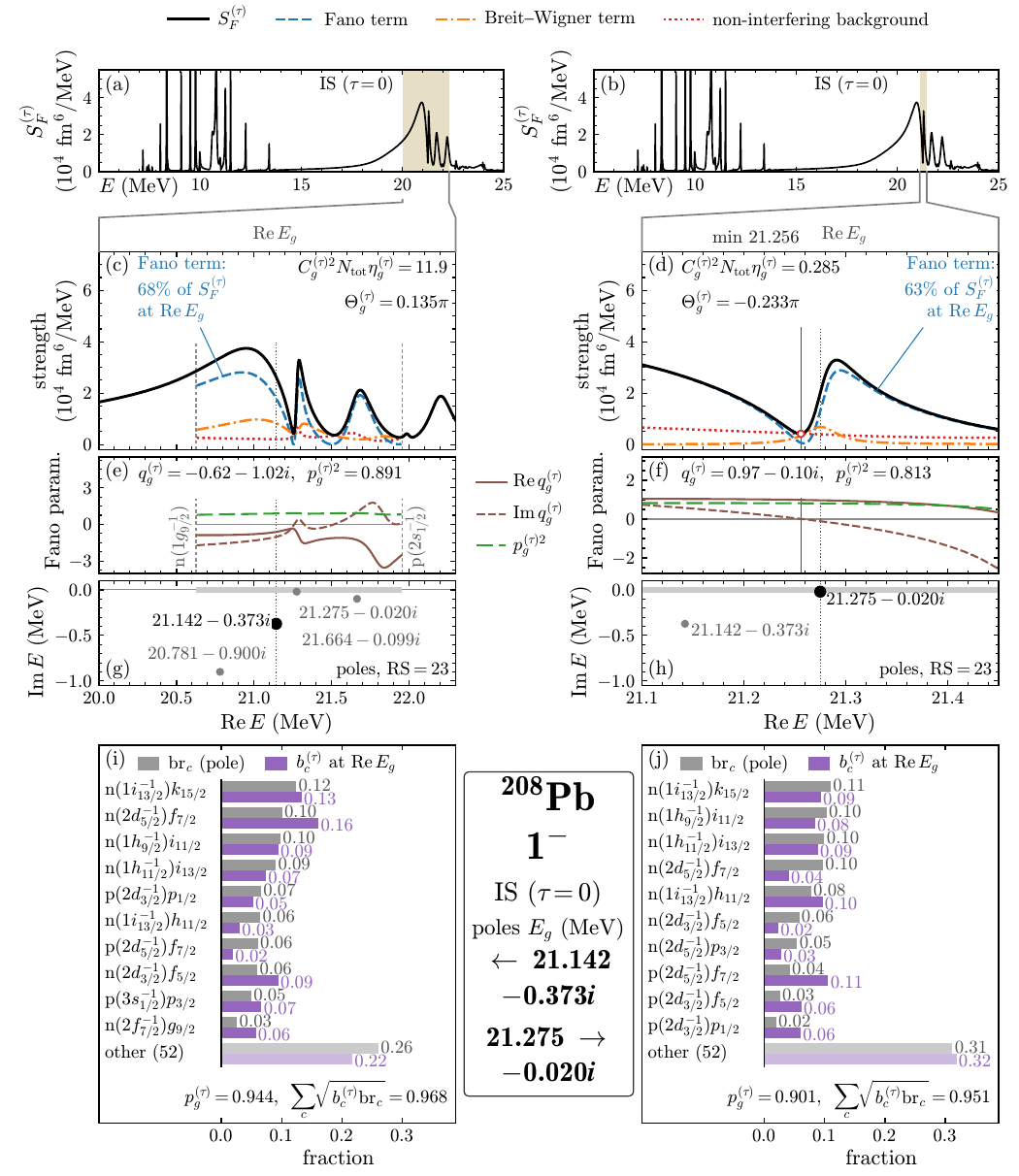}
  \caption{As Fig.~\ref{fig:dipO16}, for the isoscalar field of $^{208}$Pb $1^-$
and two poles of $\mathrm{RS}=23$,
  the ISGDR pole $21.142-0.373i$~MeV (left) and the pole $21.275-0.020i$~MeV
(right); the rows (c),(d) to (i),(j)
  correspond to Fig.~\ref{fig:dipO16}(b), (c), (d), and (e), and the labels are
as in Fig.~\ref{fig:peak15794}.
  (a),(b) Whole strength on the 10-keV grid, which does not resolve the narrow
peaks; peaks exceeding the frame are
  cut, and the band is the window. In the left column the decomposition covers
the $\mathrm{RS}=23$ interval between
  the thresholds (dashed), outside which only $S_F^{(0)}$ is shown.}
  \label{fig:pairPbIS}
\end{figure*}

\begin{figure*}[tp]
  \centering
  \includegraphics[width=\textwidth]{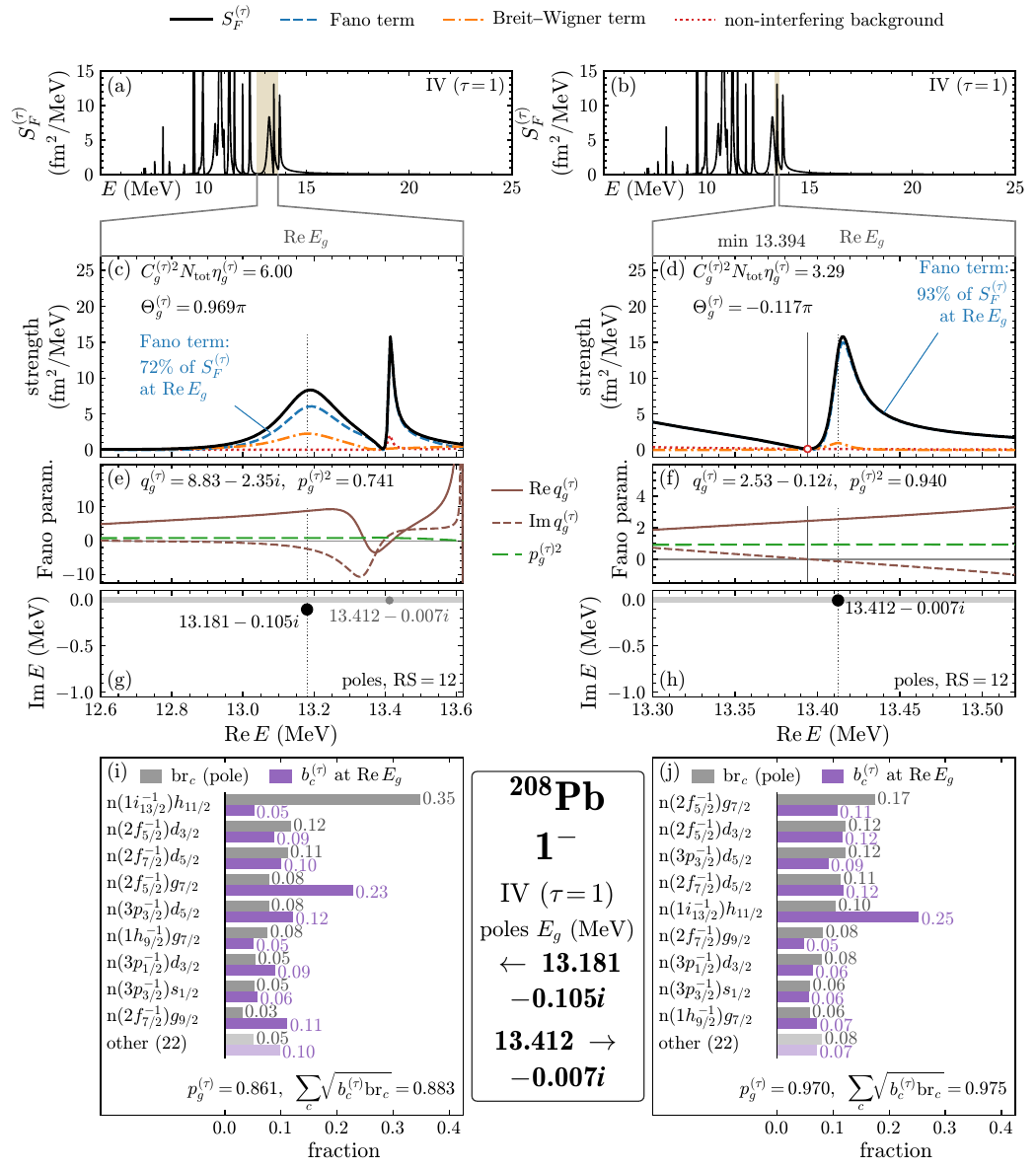}
  \caption{As Fig.~\ref{fig:pairPbIS}, for the isovector field, with the IVGDR
pole $13.181-0.105i$~MeV (left) and
  the pole $13.412-0.007i$~MeV (right) of $\mathrm{RS}=12$; here the
decomposition covers the whole window in both
  columns.}
  \label{fig:pairPbIV}
\end{figure*}

The ISGDR and the IVGDR poles have the largest indicators of
Table~\ref{tab:casesPb},
$C_g^{(0)2}\Ntot\eta_g^{(0)}=11.9$ and $C_g^{(1)2}\Ntot\eta_g^{(1)}=6.00$, and
both produce Fano-type peaks (left
columns of Figs.~\ref{fig:pairPbIS} and \ref{fig:pairPbIV}).

For the ISGDR pole in the isoscalar field, $q_g^{(0)}=-0.62-1.02i$
($|q_g^{(0)}|=1.19$) and $p_g^{(0)2}=0.891$ at
$\ReEg$, where the Fano term carries 68\% of the strength. With $|q_g^{(0)}|$
close to one and
$\operatorname{Re}q_g^{(0)}<0$, the maximum lies on the low-energy side and the
region below the background on the high-energy side. The ratio
$S_F^{(0)}/\|\vec{\mathcal{F}}^{(0)}_{\mathrm{bg}}\|^2$ reaches 2.73 at
20.874~MeV ($\epsilon=-0.72$), where
$p_g^{(0)2}=0.83$ and the Fano term carries 76\% of the strength: a Fano-type
peak. The strength falls below the background at 21.174~MeV ($\epsilon=0.09$),
and the ratio is smallest, 0.063 at 21.258~MeV, close to $1-p_g^{(0)2}=0.059$
there, at the dip of the pole at 21.275~MeV (Sec.~\ref{subsec:dipPb}), whose
term is part of the background of the ISGDR pole; near this narrow pole
$q_g^{(0)}$ changes rapidly [Fig.~\ref{fig:pairPbIS}(e)]. The branching ratios
of the ISGDR pole are spread over many channels [Fig.~\ref{fig:pairPbIS}(i)];
the bound
is 0.968, and $p_g^{(0)}=0.944$.

For the IVGDR pole in the isovector field,
$q_g^{(1)}=8.83-2.35i$ and $p_g^{(1)2}=0.741$ at $\ReEg$: a Fano-type peak with
a large $|q_g^{(1)}|$, nearly Lorentzian
near the center, where the Fano and Breit--Wigner terms carry 72\% and 28\% of
the strength. Fano's zero lies far on the
low-energy side: $S_F^{(1)}$ has its minimum at 12.665~MeV ($\epsilon=-4.9$),
0.228 of the background, close to $1-p_g^{(1)2}=0.223$ there. The strength lies
below the background from 12.85~MeV ($\epsilon=-3.1$) down to the lower end of
the window, so that the IVGDR pole also produces a dip in the sense of
Sec.~\ref{subsec:shapes}, at 1--3\% of the peak
strength. The largest branching ratio of the IVGDR pole, 0.348, is that of
$\mathrm{n}(1i_{13/2}^{-1})h_{11/2}$, in
which the isovector background is small ($b_c^{(1)}=0.052$)
[Fig.~\ref{fig:pairPbIV}(i)]; the bound is 0.883, and
$p_g^{(1)}=0.861$.

\begin{figure*}[tp]
  \centering
  \includegraphics[width=\textwidth]{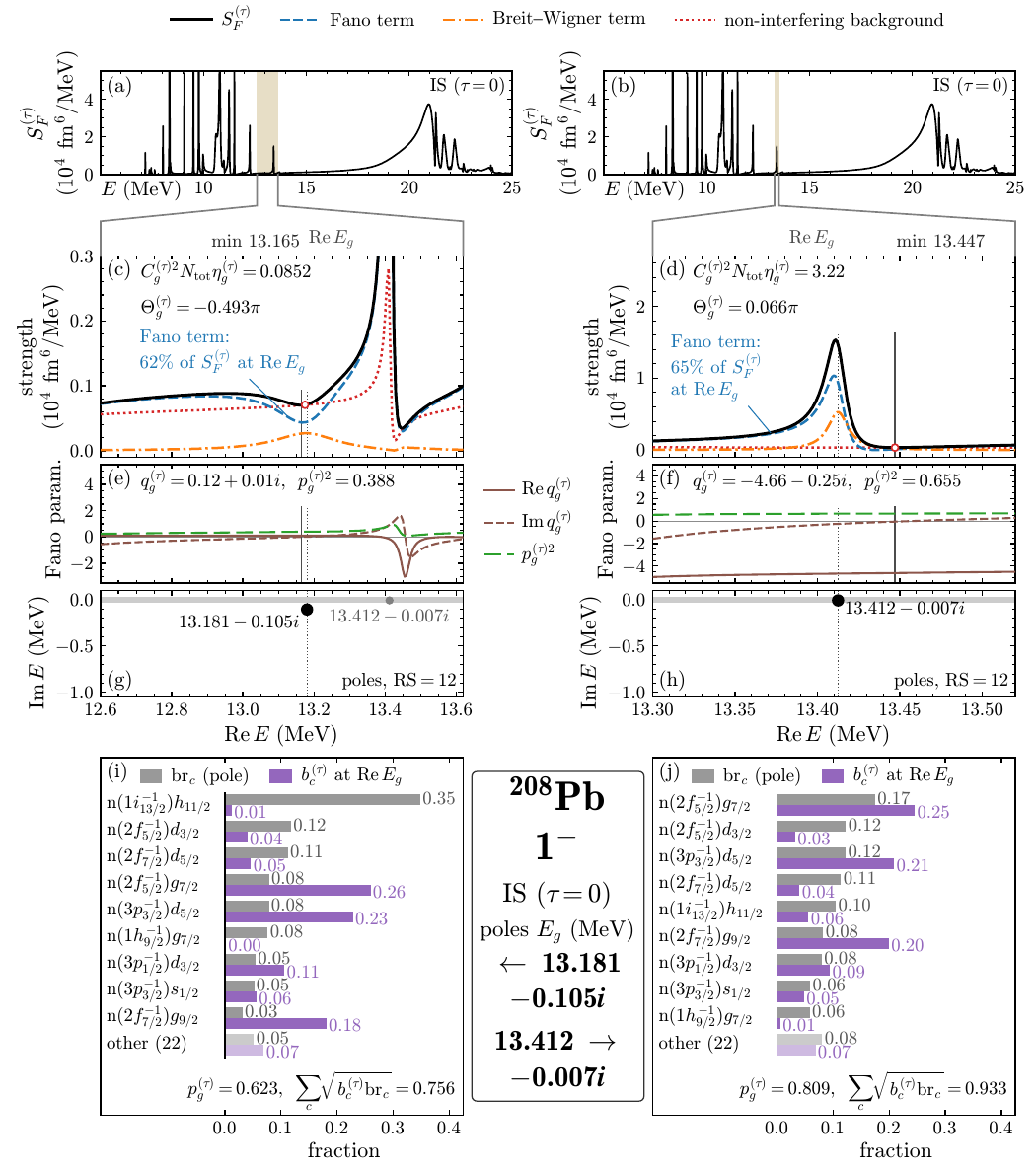}
  \caption{As Fig.~\ref{fig:pairPbIV}, for the isoscalar field; in (c) the peak
of the pole $13.412-0.007i$~MeV,
  shown in (d), is cut by the frame.}
  \label{fig:pairPbIS12}
\end{figure*}

In the isoscalar field the IVGDR pole produces a Fano dip
[Fig.~\ref{fig:pairPbIS12}, left column]. Here
$q_g^{(0)}=0.12+0.01i$ at $\ReEg$, close to the symmetric dip of $\qg=0$. The
strength at the center lies at 0.617 of the
background, close to $1-p_g^{(0)2}=0.612$, and stays below the background, which
contains the term of the pole at 13.412~MeV, over the whole window
12.60--13.62~MeV. The depth is limited by $p_g^{(0)2}=0.388$, as at the pole at
28.495~MeV of $^{16}$O.

\subsection{Narrow poles near the ISGDR and IVGDR poles}
\label{subsec:dipPb}
For the pole $21.275-0.020i$~MeV, $q_g^{(0)}=0.97-0.10i$ and $p_g^{(0)2}=0.813$
at $\ReEg$, nearly constant within the width [Fig.~\ref{fig:pairPbIS}(f)]. With
$|q_g^{(0)}|\simeq1$ the profile is Fano's asymmetric one, with its zero at
$\epsilon=-0.97$ and its maximum at $\epsilon=1.03$: a Fano-type peak with a dip
on its low-energy side. The minimum of
$S_F^{(0)}$, at 21.256~MeV ($\epsilon=-0.95$), is 0.183 of the background. At
the maximum of $S_F^{(0)}/\|\vec{\mathcal{F}}^{(0)}_{\mathrm{bg}}\|^2$, 1.75 at
21.294~MeV,
the Fano term of Eq.~(\ref{eq:fanoBW}) carries 89\% of the strength. At the
pole, Eq.~(\ref{eq:qTheta}) gives $\bar{q}_g^{(0)}=1.11-0.12i$.

Removing pole terms assigns the minimum to this pole. Without its term, the
strength decreases monotonically over the window 21.10--21.45~MeV, whereas the
minimum remains without the term of the pole $20.781-0.900i$ or
$21.664-0.099i$~MeV of the same sheet. Without the ISGDR term, the narrow
structure has the opposite asymmetry, with the minimum above the pole, at
21.358~MeV.

The branching ratios of the pole are spread over many channels, the largest
being 0.110 [Fig.~\ref{fig:pairPbIS}(j)], and the isoscalar
background at $\ReEg$ is spread over the same channels in a similar way. The
bound of Eq.~(\ref{eq:pbound}) is 0.951,
and $p_g^{(0)}=0.901$ is close to it.

In the isovector strength, the pole $13.412-0.007i$~MeV, 0.23~MeV above the
IVGDR pole, has $q_g^{(1)}=2.53-0.12i$ and
$p_g^{(1)2}=0.940$ at $\ReEg$, nearly constant within the width
[Fig.~\ref{fig:pairPbIV}(f)]. Fano's profile for
this $q$ has its zero at $\epsilon=-2.53$ and its maximum at $\epsilon=0.40$.
The minimum of $S_F^{(1)}$, at 13.394~MeV
($\epsilon=-2.47$), is 0.062 of the background. At the maximum of
$S_F^{(1)}/\|\vec{\mathcal{F}}^{(1)}_{\mathrm{bg}}\|^2$, 7.12 at 13.416~MeV, the
Fano term carries 95\% of the
strength. Since $|q_g^{(1)}|>1$, the strength lies above the background at the
center and falls below it only on the
side of the zero.

Removing pole terms again assigns the minimum to this pole. Without its term,
the isovector strength decreases monotonically over the window 13.30--13.52~MeV,
whereas the minimum remains without the term of the pole $13.697-0.019i$~MeV of
the same sheet. Without the term of the IVGDR pole, the minimum moves away from
the pole, to 13.321~MeV ($\epsilon=-12$).

The branching ratios of the pole are spread over many channels, the largest
being 0.175
[Fig.~\ref{fig:pairPbIV}(j)], and the isovector background at $\ReEg$ is spread
over the same channels; the bound is
0.975, and $p_g^{(1)}=0.970$.

The two minima are thus produced by the narrow poles, and the dips are those of
their Fano-type peaks, asymmetric Fano profiles with
$|\qg|\simeq1$ and 2.5, and the term of the ISGDR or IVGDR pole in the
background of each narrow pole places the dip close to the pole on its
low-energy side. As at the Fano
dip of $^{16}$O, at each minimum $S_F^{(\tau)}/\|\Fbg\|^2$ lies within 0.001 of
its lower bound $1-p_g^{(\tau)2}$.

In the isoscalar field, the pole at 13.412~MeV produces a mixed peak and a dip
[Fig.~\ref{fig:pairPbIS12}, right column], with
$q_g^{(0)}=-4.66-0.25i$ and $p_g^{(0)2}=0.655$ at $\ReEg$. Since
$\operatorname{Re}q_g^{(0)}<0$, the zero lies on the
high-energy side, whereas the isovector zero lies on the low-energy side. The
strength falls below the background from
13.430~MeV ($\epsilon=2.3$), and the minimum of $S_F^{(0)}$, 0.334 of the
background at 13.447~MeV, lies within
$10^{-4}$ of $1-p_g^{(0)2}$ there. This isoscalar peak and the isovector peak of
the IVGDR pole lie near the boundary
between the mixed and the Fano-type peak, with 67\% and 73\% of the strength in
the Fano term and
$p_g^{(\tau)2}=0.655$ and 0.736 at the maximum of $S_F^{(\tau)}/\|\Fbg\|^2$. In
Table~\ref{tab:casesPb} the former is
counted as mixed and the latter as a Fano-type peak.

\section{Dependence on the field, collectivity, and the phase at the pole}
\label{sec:common}

In both nuclei the same pole produces different patterns in the two fields: in
$^{16}$O the pole at 15.794~MeV and the
ISGQR (Secs.~\ref{subsec:narrow} and \ref{subsec:isgqr}), and in $^{208}$Pb the
IVGDR pole (Sec.~\ref{subsec:giant}). In both nuclei, too, a giant resonance
produces a dip in the other
field, the ISGQR in the isovector and the IVGDR pole in the isoscalar field. The
line shapes differ between the fields through $\Fbg$ and $\Fg$, while $\om$ is
common. One resonance likewise produces Breit--Wigner and Fano profiles in
different spectra in cold atom-molecule collisions~\cite{Paliwal2021} and in
dielectronic recombination~\cite{Tu2015}, and the Fano parameter of a
quasinormal mode of a split-ring resonator takes values from 0 to $\pm\infty$ as
the resonator is rotated relative to the incident wave~\cite{Bochkarev2026}.

The patterns found in Secs.~\ref{sec:O16} and \ref{sec:Pb} do not follow the
collectivity indicators. In
Tables~\ref{tab:cases} and \ref{tab:casesPb}, the indicators of the
Breit--Wigner-type peaks are 0.940 and 7.09, those of the Fano-type peaks range
from 0.0175 to 11.9, and those of the mixed peaks are 0.00769 and 3.22. Those of
the Fano dips are 0.0232 and 0.0852, and the strength also falls below the
background beside peaks with indicators up to 11.9. The IVGDR pole, which
produces a Fano dip in the isoscalar and a peak in the isovector field,
has the smaller indicator in the field of the Fano dip (0.0852 against 6.00), as
the ISGQR has in the field of its dip
(0.00769 against 7.09). The formulation does not require this, since at $E_g^*$
the ratio of the pole term to the
background along $\om$ does not depend on $|\Fg|$ (Sec.~\ref{subsec:phase}).

Fano's condition for a dip at the resonance energy, evaluated at $E_g^*$,
involves instead the phase at the pole and $\sg/\Gamma_g$, not $|\Fg|$
[Eq.~(\ref{eq:qTheta})], and both Fano dips of the tables satisfy it,
$|\bar{q}_g^{(\tau)}|<1$ (0.39 and 0.02), that is,
$\operatorname{Re}\widetilde{B}_g^{(\tau)}<0$, and for them $|\qg|<1$ holds also
at $\ReEg$, where it places the
strength at the center below the background. That no peak appears within
$|\epsilon|\le4$ is not fixed by
$\bar{q}_g^{(\tau)}$: with $q_g^{(0)}$ fixed at $\bar{q}_g^{(0)}$,
Eq.~(\ref{eq:SF-q}) changes sign at $\epsilon=1.10$ for
the pole at 28.495~MeV, and the ISGQR in the isovector field, with
$|\bar{q}_g^{(1)}|=0.70$, has a weak mixed peak beside
its dip. The peaks of the pole at 15.794~MeV in both fields, the isoscalar peak
of the ISGQR, and the isovector peak of the IVGDR pole have
$|\bar{q}_g^{(\tau)}|\ge10$. Their patterns nevertheless range from the
Breit--Wigner-type to the Fano-type peak,
because the pattern of a peak is set by $p_g^{(\tau)2}$, which the phase does
not fix. For the pole at 21.275~MeV, the
strength at the center lies just below the background although
$|\bar{q}_g^{(0)}|=1.11$, since $|q_g^{(0)}|=0.98$ at
$\ReEg$; the change of the background between $E_g^*$ and the real axis decides
whether $|q_g^{(0)}|$ at $\ReEg$ is above or below one.

\section{Summary}
\label{sec:summary}
Fano's formula has been derived for a resonance pole of the Jost-RPA strength
function. The residue of the pole term
of the transition-amplitude vector lies along the Takagi vector $\om$ for every
external field, so that $\om$ is the
combination of channels that interacts with the resonance. On the real energy
interval of the sheet of the pole, the
strength function then takes exactly Fano's form for several continuum channels,
Eq.~(\ref{eq:fano}), obtained without dividing the system into a discrete state
and a continuum. In this form $\qg$ is complex and energy dependent, and $\pg$
serves as
Fano's $|\rho|$. The second exact form, Eq.~(\ref{eq:fanoBW}), separates a Fano
term and a Breit--Wigner term. The
strength near a pole takes one of three patterns, the Breit--Wigner-type peak
for $p_g^{(\tau)2}\simeq0$, the Fano dip, and the Fano-type peak for
$p_g^{(\tau)2}\simeq1$, with mixed peaks between the two kinds of peak. The
parameter $\qg$ decides on which side of the background the strength lies, and
$p_g^{(\tau)2}$ how far; $\pg$ is bounded by the overlap of the channel
distribution of the
background with the branching ratios of the pole and is lowered further by the
relative phases. A negative real part of
the complex transition probability is equivalent to $|\bar{q}_g^{(\tau)}|<1$,
Fano's condition for a dip at the
resonance energy evaluated at $E_g^*$, and $-\cot\Theta_g^{(\tau)}$ of
Ref.~\cite{Mizuyama2026_Collectivity} is Fano's
parameter at the pole for $\sg=\Gamma_g$.

In $^{16}$O, the dips found in
Refs.~\cite{Mizuyama2025_JRPA2,Mizuyama2026_Collectivity} come from the
interference of
single poles with their backgrounds. The pole $28.495-0.564i$~MeV produces a
Fano dip in the isoscalar field, as deep as
$p_g^{(0)2}$ allows, and the ISGQR pole produces the isovector dip near its peak
energy. In both nuclei Fano dips and Fano-type peaks appear, and the same pole
produces different patterns in different fields,
because $\om$ is common to all fields whereas $\Fbg$ and $\Fg$ depend on the
field. Like the ISGQR in $^{16}$O, the IVGDR pole of $^{208}$Pb produces a dip
in the other field, here a Fano dip. The structures that the earlier Jost-RPA
analyses found in $^{16}$O are thus not specific to a light nucleus. Narrow
poles near the ISGDR and IVGDR poles produce, in the field of each resonance,
Fano-type peaks with dips that reach the lower bound of the strength. The
patterns do not follow the collectivity indicators, which enter the strength
only through $|\Fg|$, the size of the pole
term, and fix neither $\pg$ nor the phase. Fano's condition for a dip at the
resonance energy, evaluated at $E_g^*$, involves instead the phase at the pole
and $\sg/\Gamma_g$, and both Fano dips, of the pole
at 28.495~MeV of $^{16}$O and of the IVGDR pole of $^{208}$Pb, have
$|\bar{q}_g^{(\tau)}|<1$.

The analysis has limitations. The Jost-RPA includes the escape widths from the
coupling of the particle-hole configurations to the continuum but not the
spreading widths from the coupling to more complex configurations, which
semimicroscopic approaches add
phenomenologically~\cite{MuravievUrin1994,Urin2008}. The structures of the
narrow poles, with $\Gamma_g$
from 6.9 to 40~keV, are therefore properties of the RPA response; a spreading
width larger than $\Gamma_g$ would smooth
them, and an absorptive term for it would break the unitarity of $\Scc$ on which
Eq.~(\ref{eq:SF-norm}) rests. The
model, a Woods--Saxon mean field with a residual interaction of Skyrme type, is
not self-consistent, and the positions of the poles and the pattern of each pole
depend on it; no comparison with experiment is made. The Jost-RPA, however,
reproduces the strength function of the conventional continuum RPA in the
Green-function method~\cite{ShlomoBertsch1975}, with which it was compared for
$^{16}$O in Ref.~\cite{Mizuyama2023_JRPA1}. The values of $\qg$ and $\pg$ at
$\ReEg$ apply across the width only when the background changes slowly, and each
decomposition is restricted to the real interval of one Riemann sheet. The
decomposition and the patterns are defined with the free waves
$rh^{(\pm)}_l(kr)$ to which the Jost solutions reduce for $r\ge R_{\max}$
(Appendix~\ref{subsec:Smatrix}). By the multichannel form of Watson's
final-state relation, $\Fvec=\Scc\vec{\mathcal{F}}^{(\tau)*}$ on the real
axis~\cite{Watson1952,Watson1954}, the phase of $\Fvec$ is tied to that of
$\Scc$; an energy-dependent rephasing of the scattering solutions multiplies
$\Fvec$ by the phase factors and $\Scc$ by them on both sides, which leaves
$S_F^{(\tau)}$, $E_g$, and $\widetilde{B}_g^{(\tau)}$, with the sign of its real
part, unchanged but in general changes $\om$, $\sg$, $\mathrm{br}_c$,
$\bar{q}_g^{(\tau)}$, $\Fbg$, $\qg$, and $\pg$, and with them the patterns.

The derivation uses three properties of the Jost-RPA. First, the strength
function is the squared norm of an amplitude vector in the open channels, which
rests on the symmetry and unitarity of $\Scc$ on the real axis. Second, the
residue of $\Scc$ at a non-degenerate pole, a simple zero of
$\det\mathcal{J}^{(+)}(E)$, is symmetric and of rank one. Third, the vectors
$\vec{\rho}^{\,(-)}_q$ are analytic at the pole. The two forms, the bound on
$\pg$, and the three patterns hold in any
framework with these properties, and the relation to the phase at the pole
requires in addition that the response
function be analytic in the upper half plane of the physical sheet. None of
these properties is specific to nuclei, so that the two forms, with a
Breit--Wigner term that requires no decoherence, apply also to other open
quantum systems coupled to the continuum in which these properties hold, such as
atomic, molecular, and mesoscopic systems with several open channels. The
formulation also applies to the quasiparticle resonances of the
Hartree--Fock--Bogoliubov theory formulated within the Jost function framework,
for which the Takagi factorization was introduced~\cite{Mizuyama2026_Takagi},
once the strength function is written as the squared norm of an amplitude vector
in the open quasiparticle channels; a reanalysis of the resonances of
Ref.~\cite{Mizuyama2020_FanoHFB} with the Fano parameter defined by the pole is
left for future work.

\appendix
\section{Jost-RPA framework and model}
\label{app:jrpa}
This appendix gives the model, the numerical details, the derivations of the
results summarized in Sec.~\ref{sec:formalism}, and the normalization
convention. Most of these results were obtained in
Refs.~\cite{Mizuyama2023_JRPA1,Mizuyama2024_Eigenphase,Mizuyama2025_JRPA2,%
Mizuyama2026_Takagi,Mizuyama2026_Collectivity}.

\subsection{Model, channel space, and numerical details}
\label{subsec:model}
The channels $i=1,\dots,\Ntot$ are those of Sec.~\ref{sec:formalism}. Forward
and backward p-h amplitudes are treated together, so that the radial wave
functions are $2\Ntot$-dimensional vectors, and the regular and irregular
solutions are $2\Ntot\times2\Ntot$
matrices. At the excitation energy $E$, the forward and backward components of
channel $i$ propagate with the wave
numbers
\begin{align}
  k_1^{(i)}(E)&=\sqrt{\frac{2m}{\hbar^2}\bigl(\epsilon_{h_i}+E\bigr)},
  \nonumber\\
  k_2^{(i)}(E)&=\sqrt{\frac{2m}{\hbar^2}\bigl(\epsilon_{h_i}-E\bigr)},
  \label{eq:k}
\end{align}
which form the diagonal matrix
$\mathcal{K}(E)=\operatorname{diag}(k_1^{(1)},\dots,k_1^{(\Ntot)},
k_2^{(1)},\dots,k_2^{(\Ntot)})$. The regular solution $\Phi^{(r)}(r;E)$ and the
irregular (Jost) solutions
$\Phi^{(\pm)}(r;E)$ satisfy the coupled radial equations
\begin{equation}
  \left[\mathbf{I}\frac{d^2}{dr^2}-\frac{\mathbf{L}^2}{r^2}+\mathcal{K}^2(E)
  -\frac{2m}{\hbar^2}\mathcal{V}(r)\right]\Phi(r;E)=0,
  \label{eq:radial}
\end{equation}
where
$\mathbf{L}^2=\operatorname{diag}\bigl(l_1(l_1+1),\dots,l_{\Ntot}(l_{\Ntot}+1),%
l_1(l_1+1),\dots,
l_{\Ntot}(l_{\Ntot}+1)\bigr)$ and
\begin{equation}
  \mathcal{V}(r)=
  \begin{pmatrix}
    \mathcal{U}(r)+\mathcal{V}_{\mathrm{res}}(r) &
\mathcal{V}_{\mathrm{res}}(r)\\
    \mathcal{V}_{\mathrm{res}}(r) & \mathcal{U}(r)+\mathcal{V}_{\mathrm{res}}(r)
  \end{pmatrix}.
  \label{eq:Vmat}
\end{equation}
The diagonal matrix $\mathcal{U}(r)=\operatorname{diag}(U_1,\dots,U_{\Ntot})$
contains the mean-field potential
of the particle in each channel,
\begin{align}
  U_i(r)={}&\Bigl(1-0.67\,\tau_q\frac{N-Z}{A}\Bigr)
  \nonumber\\
  &\times\Bigl[V_0\,f_{\mathrm{WS}}\Bigl(\frac{r-R}{a}\Bigr)
  -\frac{V_{\mathrm{ls}}}{ra}\,\langle\mathbf{l}\cdot\bm{\sigma}\rangle_i\,
  g_{\mathrm{WS}}\Bigl(\frac{r-R}{a}\Bigr)\Bigr]
  \nonumber\\
  &+\delta_{qp}V_{\mathrm{C}}(r),
  \label{eq:WS}
\end{align}
with $\tau_n=+1$, $\tau_p=-1$,
$\langle\mathbf{l}\cdot\bm{\sigma}\rangle=j(j+1)-l(l+1)-3/4$,
$f_{\mathrm{WS}}(x)=1/(1+e^x)$, $g_{\mathrm{WS}}(x)=e^x/(1+e^x)^2$, the radius
$R=r_0(A-1)^{1/3}$, and the
Coulomb potential $V_{\mathrm{C}}(r)$ of a uniformly charged sphere of charge
$Ze$ and radius $R$, which is cut at
the matching radius $R_{\max}$ (see below). The same potential, with the
angular momenta of the hole, gives the hole wave functions $\varphi_{h}(r)$ and
the energies $\epsilon_h$. The
residual interaction is a density-dependent contact interaction of Skyrme type,
\begin{equation}
  \mathcal{V}_{\mathrm{res}}(r)=\sum_{q,q'}\tilde{\varphi}_q(r)\,\frac{%
\kappa_{qq'}(r)}{r^2}\,
  \tilde{\varphi}_{q'}^{\Tr}(r),
  \label{eq:Vres}
\end{equation}
with
\begin{align}
  \kappa_{qq}(r)&=\tfrac{1}{2}t_0(1-x_0)
  \nonumber\\
  &\quad+\tfrac{1}{12}t_3\bigl[(5+x_3)\rho(r)-(2+4x_3)\rho_q(r)\bigr],
  \label{eq:kqq}\\
  \kappa_{qq'}(r)&=t_0\bigl(1+\tfrac{1}{2}x_0\bigr)+\tfrac{1}{12}t_3(5+x_3)%
\rho(r),
  \label{eq:kqqbar}
\end{align}
where $\rho_q(r)$ is the ground-state density of the nucleons $q$,
$\rho=\rho_n+\rho_p$, and $q'\neq q$ in Eq.~(\ref{eq:kqqbar}).
The $\Ntot$-dimensional vector $\tilde{\varphi}_q(r)$ has the components
\begin{equation}
  \tilde{\varphi}_{q,i}(r)=\frac{\langle
l_ij_i\|Y_L\|l_{h_i}j_{h_i}\rangle}{\sqrt{2L+1}}\,\varphi_{h_i}(r)
  \label{eq:phitilde}
\end{equation}
for the channels $i$ of the nucleons $q$, and zero for the other channels.

The parameters of the mean field are $V_0=-60$~MeV,
$V_{\mathrm{ls}}=15$~MeV~fm$^2$, $r_0=1.2$~fm, and
$a=0.65$~fm for $^{16}$O, as in the earlier Jost-RPA
studies~\cite{Mizuyama2023_JRPA1,Mizuyama2024_Eigenphase,Mizuyama2025_JRPA2,%
Mizuyama2026_Collectivity}. For
$^{208}$Pb, $V_0=-51.0$~MeV, $r_0=1.27$~fm, and $a=0.67$~fm are taken from
Ref.~\cite{BohrMottelson1969} and used
with the radius and the isospin factor of Eq.~(\ref{eq:WS}). The value
$V_{\mathrm{ls}}=0.44|V_0|r_0^2/2=18.10$~MeV~fm$^2$
for $^{208}$Pb corresponds to the spin-orbit strength of
Ref.~\cite{BohrMottelson1969} (the factor 1/2 because Eq.~(\ref{eq:WS}) is
written with $\mathbf{l}\cdot\bm{\sigma}$ instead of
$\mathbf{l}\cdot\mathbf{s}$).
The residual interaction uses $t_0=-1100$~MeV~fm$^3$, $t_3=16000$~MeV~fm$^6$,
$x_0=0.5$, and $x_3=1$ for both nuclei. The radial equations are
solved up to $R_{\max}=20$~fm with the renormalized Numerov
method~\cite{Johnson1977} in its coupled-channel
form~\cite{Johnson1978} on a mesh of 100 points for
$^{16}$O and 80 points for $^{208}$Pb. The Coulomb potential is cut at
$R_{\max}$, where the proton channels are
matched to free spherical Hankel functions; the cut produces a few poles of
exterior proton waves with vanishing
residues, which are discarded. The poles are located with the contour-integral
root-finding method for
analytic functions~\cite{Kravanja2000}, on which the contour-integral eigenvalue
method of
Ref.~\cite{Sakurai2003} is based. The search is made in a rectangle whose
real-axis edge is the energy interval of the
sheet and whose depth is $-2~\mathrm{MeV}\le\operatorname{Im}E<0$ for $^{16}$O
and $-1~\mathrm{MeV}\le\operatorname{Im}E<0$
for $^{208}$Pb. For $^{16}$O all sheets $\mathrm{RS}=2$--$7$ are searched (on
the last one up to 50~MeV); for
$^{208}$Pb the sheets $\mathrm{RS}=2$--$12$ and $21$--$25$, which cover
$7.04$--$14.61$~MeV and $17.94$--$23.07$~MeV.

The amplitude $\Fvec(E)$ of each pole analyzed is computed on the real interval
of its sheet or on a part of it. For
$^{16}$O the grid is 10~keV for the poles $27.038-1.456i$, $28.495-0.564i$, and
$30.067-0.660i$~MeV (26--31~MeV) and
1~keV for the other poles (12.00--16.43~MeV for $15.794-0.003i$~MeV and the
interval 16.43--16.89~MeV of
$\mathrm{RS}=3$ for the ISGQR). For $^{208}$Pb it is 4~keV for the pole
$21.142-0.373i$~MeV
(20.63--21.95~MeV), 2~keV for $21.275-0.020i$~MeV (21.10--21.45~MeV), 5~keV for
$13.181-0.105i$~MeV
(12.60--13.62~MeV), and 1~keV for $13.412-0.007i$~MeV (13.30--13.52~MeV). The
heights given in the figures for peaks that exceed their frames are the maxima
on the grids of the decompositions, except that of the peak at 17.136~MeV in
Fig.~\ref{fig:peak15794}(a), which is the maximum on a 1-keV grid over
16.89--21.34~MeV.

\subsection{Jost functions and the open-channel $S$ matrix}
\label{subsec:Smatrix}
The regular solution behaves as
$\Phi^{(r)}_{\alpha\beta}(r;E)\to\delta_{\alpha\beta}\,r\,j_{l_\alpha}(k_\alpha
r)$
for $r\to0$, and the Jost solutions as
$\Phi^{(\pm)}_{\alpha\beta}(r;E)=\delta_{\alpha\beta}\,r\,
h^{(\pm)}_{l_\alpha}(k_\alpha r)$ for $r\ge R_{\max}$, where $j_l$ is the
spherical Bessel function and
$h^{(\pm)}_l$ are the outgoing and incoming spherical Hankel functions. The Jost
matrices
$\mathcal{J}^{(\pm)}(E)$ connect the two sets of solutions,
\begin{align}
  \Phi^{(r)\Tr}(r;E)=\frac{1}{2}\Bigl[&\mathcal{J}^{(+)}(E)\Phi^{(-)\Tr}(r;E)
  \nonumber\\
  &+\mathcal{J}^{(-)}(E)\Phi^{(+)\Tr}(r;E)\Bigr],
  \label{eq:Jost}
\end{align}
and are evaluated from the matrix Wronskian of $\Phi^{(r)}$ and
$\Phi^{(\pm)}$~\cite{Mizuyama2023_JRPA1}. The
flux-adjusted $S$ matrix,
\begin{equation}
  \mathcal{S}(E)=\mathcal{K}^{\frac{1}{2}}\bigl[\mathcal{J}^{(+)}(E)\bigr]^{-1}%
\mathcal{J}^{(-)}(E)\,
  \mathcal{K}^{-\frac{1}{2}},
  \label{eq:Smat}
\end{equation}
is complex symmetric, $\mathcal{S}^{\Tr}=\mathcal{S}$, as a consequence of the
reciprocity of the Jost
matrices~\cite{Mizuyama2024_Eigenphase,Mizuyama2026_Collectivity}. Its poles
$E_g=\operatorname{Re}E_g-i\Gamma_g/2$ are the zeros of
$\det\mathcal{J}^{(+)}(E)$. Partitioning the channels into the open ($c$) and
closed ($d$) ones,
\begin{equation}
  \mathcal{S}=
  \begin{pmatrix}
    \mathcal{S}_{cc} & \mathcal{S}_{cd}\\
    \mathcal{S}_{dc} & \mathcal{S}_{dd}
  \end{pmatrix},
  \label{eq:Spart}
\end{equation}
the open-channel block $\Scc$ is unitary on the real energy interval of the
sheet~\cite{Mizuyama2024_Eigenphase},
\begin{equation}
  \Scc(E)\,\Scc^{\dagger}(E)=\mathbf{1},
  \label{eq:unitarity}
\end{equation}
because the closed channels decay exponentially and carry no asymptotic flux.
Since
$\Scc$ is symmetric, Eq.~(\ref{eq:unitarity}) reads $\Scc^{-1}(E)=[\Scc(E)]^{*}$
on the real axis, and its
analytic continuation,
\begin{equation}
  \Scc^{-1}(E)=\bigl[\Scc(E^*)\bigr]^{*},
  \label{eq:unitarity-cont}
\end{equation}
is analytic at $E=E_g$, since $E_g^*$ lies in the upper half plane of the
physical sheet, where $\Scc$ has no
pole~\cite{Mizuyama2026_Collectivity}.

\subsection{Strength function and the open-channel $S$ matrix}
\label{subsec:strength}
The Jost-RPA Green function, the solution of
\begin{align}
  \Bigl[\frac{\hbar^2}{2m}\mathcal{K}^2&-\Bigl\{-\frac{\hbar^2}{2m}\Bigl(%
\frac{d^2}{dr^2}\mathbf{I}
  -\frac{\mathbf{L}^2}{r^2}\Bigr)+\mathcal{V}(r)\Bigr\}\Bigr]
  \nonumber\\
  &\times\mathcal{G}(r,r';E)=\mathbf{I}\,\delta(r-r'),
  \label{eq:Green-eq}
\end{align}
with outgoing boundary conditions, is
\begin{align}
  \mathcal{G}(r,r';E)={}&-i\frac{2m}{\hbar^2}\Bigl[\theta(r-r')\,\Phi^{(+)}%
_F(r)\,\Psi^{(+)\Tr}_F(r')
  \nonumber\\
  &+\theta(r'-r)\,\Psi^{(+)}_F(r)\,\Phi^{(+)\Tr}_F(r')\Bigr],
  \label{eq:Green}
\end{align}
where $\Phi^{(\pm)\Tr}_F\equiv\mathcal{K}^{1/2}\Phi^{(\pm)\Tr}$, and the
scattering solution
\begin{align}
  \Psi^{(+)\Tr}_F(r;E)&\equiv\mathcal{K}^{\frac{1}{2}}\bigl[\mathcal{J}^{(+)}%
(E)\bigr]^{-1}\Phi^{(r)\Tr}(r;E)
  \nonumber\\
  &=\frac{1}{2}\Bigl[\Phi^{(-)\Tr}_F(r;E)+\mathcal{S}(E)\,\Phi^{(+)\Tr}_F(r;E)%
\Bigr]
  \label{eq:Psi}
\end{align}
contains the outgoing and incoming waves weighted by the $S$
matrix~\cite{Mizuyama2024_Eigenphase}. With the
$2\Ntot$-dimensional hole vectors
$\xi_q(r)\equiv(\tilde{\varphi}_q(r),\tilde{\varphi}_q(r))^{\Tr}$, the RPA
response function for the external field $f^{(\tau)}$ is
\begin{align}
  R_F^{(\tau)}(E)=\sum_{q,q'}\int_0^\infty\!dr\int_0^\infty\!dr'\,&f_q^{(\tau)}%
(r)\,\xi_q^{\Tr}(r)\,
  \mathcal{G}(r,r';E)
  \nonumber\\
  &\times\xi_{q'}(r')\,f_{q'}^{(\tau)}(r'),
  \label{eq:RF}
\end{align}
and the strength function on the real axis is
\begin{equation}
  S_F^{(\tau)}(E)=-\frac{1}{\pi}\operatorname{Im}R_F^{(\tau)}(E).
  \label{eq:SF}
\end{equation}
The isoscalar ($\tau=0$) and isovector ($\tau=1$) fields are $f_q^{(0)}=r^L$ and
$f_q^{(1)}=\tau_q r^L$ for
$L\ge2$; for $L=1$, $f_q^{(0)}=r^3-\tfrac{5}{3}\langle r^2\rangle r$, with the
mean-square radius $\langle r^2\rangle$ of the
ground-state density $\rho$, and $f_n^{(1)}=(Z/A)r$, $f_p^{(1)}=-(N/A)r$.

Only the $n_c$ open-channel scattering solutions enter the strength function.
With the solutions continued to complex energies,
$\Psi^{(-)}_{F,c}(r;E)\equiv\Psi^{(+)*}_{F,c}(r;E^*)$, where $\Psi_{F,c}$
denotes the $n_c$ columns of $\Psi_F$
that belong to the open channels, the open-channel scattering solutions satisfy
$\Psi^{(+)\Tr}_{F,c}(r;E)=\Scc(E)\Psi^{(-)\Tr}_{F,c}(r;E)$~%
\cite{Mizuyama2024_Eigenphase}. The imaginary part of
the Green function, continued to complex energies as
$[\mathcal{G}(E)-\mathcal{G}^*(E^*)]/2i$, reduces
to~\cite{Mizuyama2024_Eigenphase}
\begin{align}
  &\frac{1}{2i}\Bigl[\mathcal{G}(r,r';E)-\mathcal{G}^{*}(r,r';E^*)\Bigr]
  \nonumber\\
  &\quad=-\frac{2m}{\hbar^2}\,\Psi^{(-)}_{F,c}(r;E)\,\Scc(E)\,\Psi^{(-)\Tr}_{F,%
c}(r';E).
  \label{eq:ImG}
\end{align}
Inserting
Eq.~(\ref{eq:ImG}) into Eqs.~(\ref{eq:RF}) and (\ref{eq:SF}) gives the strength
function as a bilinear form of
$\Scc$,
\begin{align}
  S_F^{(\tau)}(E)={}&\Bigl(\sum_q\int_0^\infty\!dr\,f_q^{(\tau)}(r)\,\vec{\rho}%
^{\,(-)\Tr}_q(r;E)\Bigr)
  \,\Scc(E)
  \nonumber\\
  &\times\Bigl(\sum_{q'}\int_0^\infty\!dr'\,\vec{\rho}^{\,(-)}_{q'}(r';E)\,%
f_{q'}^{(\tau)}(r')\Bigr),
  \label{eq:SF-bilinear}
\end{align}
with the $n_c$-dimensional vectors
$\vec{\rho}^{\,(-)}_q(r;E)$ of Sec.~\ref{subsec:SFnorm}. Since
$\vec{\mathcal{F}}^{(\tau)\dagger}(E^*)=\sum_q\int_0^\infty\!dr\,f_q^{(\tau)}%
(r)\,\vec{\rho}^{\,(-)\Tr}_q(r;E)$, which follows from
$\Psi^{(+)\Tr}_{F,c}=\Scc\Psi^{(-)\Tr}_{F,c}$, the reality of $\xi_q$ and
$f_q^{(\tau)}$, the symmetry of $\Scc$, and Eq.~(\ref{eq:unitarity-cont}), the
strength function is the squared norm of the transition-amplitude vector in the
open channels, Eqs.~(\ref{eq:SF-norm}) and (\ref{eq:Fvec}).

\subsection{Residue of the Green function}
\label{subsec:takagi}
At a non-degenerate pole, a simple zero of $\det\mathcal{J}^{(+)}(E)$, the
residue of $\Scc$ has the Takagi
form~(\ref{eq:takagi})~\cite{Autonne1915,Takagi1924}, since
$[\mathcal{J}^{(+)}]^{-1}=\operatorname{adj}\mathcal{J}^{(+)}/\det\mathcal{J}%
^{(+)}$ and the adjugate of
$\mathcal{J}^{(+)}(E_g)$ has rank one at a simple zero of the determinant.
Inserting Eq.~(\ref{eq:takagi}) into Eq.~(\ref{eq:ImG}), in which
$\mathcal{G}^{*}(r,r';E^*)$ and $\Psi^{(-)}_{F,c}(r;E)$ are analytic at $E_g$,
gives the residue of the Green function at the
pole~\cite{Mizuyama2026_Collectivity},
\begin{equation}
  \Res\bigl[\mathcal{G}(r,r'),E_g\bigr]=\vec{\psi}_g(r)\,\vec{\psi}_g^{\Tr}(r'),
  \label{eq:resG}
\end{equation}
with $\vec{\psi}_g$ of Eq.~(\ref{eq:psig}). With the transition
density~(\ref{eq:trd}), the residue of the response
function~(\ref{eq:RF}) is Eq.~(\ref{eq:Btilde}) with the
amplitudes~(\ref{eq:Famp}).

\subsection{Branching ratios and the residue width}
\label{subsec:branching}
Separating the pole term of $g$ from $\Scc$,
\begin{equation}
  \Scc(E)=\mathcal{S}_{\mathrm{B},g}(E)+\frac{\sg\,\om\,\vec{\omega}_g^{(c)\Tr}%
}{E-E_g},
  \label{eq:Sdecomp}
\end{equation}
where $\mathcal{S}_{\mathrm{B},g}$ contains everything except the pole term of
$g$, including the other poles, and
is analytic at $E_g$ and $E_g^*$. Because $\Scc^{-1}\Scc=\mathbf{1}$ has no pole
at $E_g$ and $\Scc^{-1}$ is
analytic there [Eq.~(\ref{eq:unitarity-cont})], $\Scc^{-1}(E_g)\om=0$, and the
complex conjugate of this relation
gives $\Scc(E_g^*)\vec{\omega}_g^{(c)*}=0$. Inserting Eq.~(\ref{eq:Sdecomp}) at
$E=E_g^*$, where
$E_g^*-E_g=i\Gamma_g$, gives~\cite{Mizuyama2026_Collectivity}
\begin{align}
  \mathcal{S}_{\mathrm{B},g}(E_g^*)\,\vec{\omega}_g^{(c)*}&=i\,\frac{\sg}{%
\Gamma_g}\,\om,
  \nonumber\\
  \frac{\sg}{\Gamma_g}&=\bigl\|\mathcal{S}_{\mathrm{B},g}(E_g^*)\,\vec{\omega}%
_g^{(c)*}\bigr\|.
  \label{eq:sigma-gamma}
\end{align}
For an energy-independent $\mathcal{S}_{\mathrm{B},g}$, for which
$\sg=\Gamma_g$, the first line has the form of the relation
between the direct-scattering matrix and the coupling vector in the temporal
coupled-mode theory of Fano resonances~\cite{Fan2003}.
Evaluating Eq.~(\ref{eq:Sdecomp}) at $E=\operatorname{Re}E_g$, expanding
$\mathcal{S}_{\mathrm{B},g}$ to first
order over the distance $\Gamma_g/2$ between $\operatorname{Re}E_g$ and $E_g^*$,
and using the
unitarity~(\ref{eq:unitarity}) of $\Scc(\operatorname{Re}E_g)$
gives~\cite{Mizuyama2026_Collectivity}
\begin{equation}
  \frac{\sg}{\Gamma_g}\simeq1-\frac{\Gamma_g}{2}\operatorname{Re}\Bigl[\vec{%
\omega}_g^{(c)\dagger}
  \frac{d\mathcal{S}_{\mathrm{B},g}}{dE}(\operatorname{Re}E_g)\,\vec{\omega}%
_g^{(c)*}\Bigr].
  \label{eq:sigma-gamma-first}
\end{equation}
Hence $\sg\simeq\Gamma_g$ for a narrow pole across which
$\mathcal{S}_{\mathrm{B},g}$ varies little, and deviations
of either sign arise from the energy dependence of $\mathcal{S}_{\mathrm{B},g}$
along the Takagi vector, caused by
thresholds, barriers, and neighboring poles. For the poles analyzed in
Secs.~\ref{sec:O16} and \ref{sec:Pb}, $\sg/\Gamma_g$ lies between 1.00 and 1.04,
except 1.10 for the pole $28.495-0.564i$~MeV, 1.69 for the ISGDR pole, and 1.12
for the pole $21.275-0.020i$~MeV, which lies within the width of the ISGDR pole.

\subsection{Normalization convention}
\label{subsec:normalization}
In the present paper the response function~(\ref{eq:RF}) is defined without a
factor $1/\pi$, so that
$S_F^{(\tau)}=-\operatorname{Im}R_F^{(\tau)}/\pi$ and
$\widetilde{B}_g^{(\tau)}=\Res[R_F^{(\tau)},E_g]=\mathcal{F}_g^{(\tau)2}$.
In Ref.~\cite{Mizuyama2026_Collectivity} the response function includes the
factor $1/\pi$ and the complex
transition probability is defined as $\pi$ times its residue; the two
definitions give the same
$\widetilde{B}_g^{(\tau)}$. The wave function $\vec{\psi}_g$, the transition
densities, and the transition amplitudes
of the present paper are $\sqrt{\pi}$ times those of
Ref.~\cite{Mizuyama2026_Collectivity}. The quantities
$C_g^{(\tau)}$, $\eta_g^{(\tau)}$, $C_g^{(\tau)2}\Ntot\eta_g^{(\tau)}$,
$\Theta_g^{(\tau)}$, and
$\mathrm{br}_c$ are not affected.

\section{Complex transition probability and Fano's parameter at the pole}
\label{app:phase}
On the real axis, Eqs.~(\ref{eq:SF}) and (\ref{eq:SF-norm}) are two forms of the
strength function. Continued to complex
energies, as in Eq.~(\ref{eq:ImG}), they read
\begin{equation}
  -\frac{1}{2\pi i}\Bigl[R_F^{(\tau)}(E)-R_F^{(\tau)*}(E^*)\Bigr]
  =\vec{\mathcal{F}}^{(\tau)\dagger}(E^*)\,\Fvec(E).
  \label{eq:SF-cont}
\end{equation}
Both sides are continued analytically from the real interval of the sheet, where
they coincide, so that they coincide
also at complex $E$ near $E_g$. The functions $R_F^{(\tau)*}(E^*)$ and
$\vec{\mathcal{F}}^{(\tau)\dagger}(E^*)$ are
analytic at $E=E_g$, because $E_g^*$ lies in the upper half plane of the
physical sheet, where $R_F^{(\tau)}$ and
$\Fvec$ are analytic. The pole at $E_g$ therefore comes from $R_F^{(\tau)}(E)$
on the left side, with the residue
$(i/2\pi)\widetilde{B}_g^{(\tau)}$ by Eq.~(\ref{eq:Btilde}), and from $\Fvec(E)$
on the right side, with the residue
$\vec{\mathcal{F}}^{(\tau)\dagger}(E_g^*)\,\cg\om=\cg\,[\omd\Fvec(E_g^*)]^*$ by
Eq.~(\ref{eq:resF}). At
$E=E_g^*$, that is, at $\epsilon=i$, the first line of Eq.~(\ref{eq:qg}) gives
\begin{align}
  \omd\Fvec(E_g^*)&=\omd\Fbg(E_g^*)\,\frac{\bar{q}_g^{(\tau)}+i}{2i}
  \nonumber\\
  &=-\frac{i\cg}{\Gamma_g}\,\frac{\bar{q}_g^{(\tau)}+i}{\bar{q}_g^{(\tau)}-i},
  \label{eq:resF-Egstar}
\end{align}
where the second form uses Eq.~(\ref{eq:qg}) at $E_g^*$. Equating the two
residues and using
$|\cg|^2=\sg|\Fg|^2/2\pi$ [Eq.~(\ref{eq:c2coll})] gives Eq.~(\ref{eq:Bres}).


\begin{thebibliography}{99}

\bibitem{Fano1961}
U. Fano,
Phys. Rev. \textbf{124}, 1866 (1961).

\bibitem{FanoCooper1965}
U. Fano and J. W. Cooper,
Phys. Rev. \textbf{137}, A1364 (1965).

\bibitem{Miroshnichenko2010}
A. E. Miroshnichenko, S. Flach, and Y. S. Kivshar,
Rev. Mod. Phys. \textbf{82}, 2257 (2010).

\bibitem{Limonov2017}
M. F. Limonov, M. V. Rybin, A. N. Poddubny, and Y. S. Kivshar,
Nat. Photonics \textbf{11}, 543 (2017).

\bibitem{SoneHyodo2025}
K. Sone and T. Hyodo,
Phys. Rev. C \textbf{112}, 065207 (2025).

\bibitem{Okolowicz2003}
J. Oko{\l}owicz, M. P{\l}oszajczak, and I. Rotter,
Phys. Rep. \textbf{374}, 271 (2003).

\bibitem{Kleinwachter1986}
P. Kleinw\"achter and I. Rotter,
J. Phys. G \textbf{12}, 821 (1986).

\bibitem{Sokolov1997a}
V. V. Sokolov, I. Rotter, D. V. Savin, and M. M\"uller,
Phys. Rev. C \textbf{56}, 1031 (1997).

\bibitem{Sokolov1997b}
V. V. Sokolov, I. Rotter, D. V. Savin, and M. M\"uller,
Phys. Rev. C \textbf{56}, 1044 (1997).

\bibitem{AuerbachZelevinsky2011}
N. Auerbach and V. Zelevinsky,
Rep. Prog. Phys. \textbf{74}, 106301 (2011).

\bibitem{Orrigo2006}
S. E. A. Orrigo, H. Lenske, F. Cappuzzello, A. Cunsolo, A. Foti, A. Lazzaro, C.
Nociforo, and J. S. Winfield,
Phys. Lett. B \textbf{633}, 469 (2006).

\bibitem{Mizuyama2020_FanoHFB}
K. Mizuyama, N. Nhu Le, and T. V. Nhan Hao,
Phys. Rev. C \textbf{101}, 034601 (2020).

\bibitem{RacecWulf2001}
E. R. Racec and U. Wulf,
Phys. Rev. B \textbf{64}, 115318 (2001).

\bibitem{Racec2010}
E. R. Racec, U. Wulf, and P. N. Racec,
Phys. Rev. B \textbf{82}, 085313 (2010).

\bibitem{Mendoza2008}
M. Mendoza, P. A. Schulz, R. O. Vallejos, and C. H. Lewenkopf,
Phys. Rev. B \textbf{77}, 155307 (2008).

\bibitem{Mies1968}
F. H. Mies,
Phys. Rev. \textbf{175}, 164 (1968).

\bibitem{Magunov2003}
A. I. Magunov, I. Rotter, and S. I. Strakhova,
Phys. Rev. B \textbf{68}, 245305 (2003).

\bibitem{Sasada2011}
K. Sasada, N. Hatano, and G. Ordonez,
J. Phys. Soc. Jpn. \textbf{80}, 104707 (2011).

\bibitem{Alvarez1989}
G. Alvarez and H. J. Silverstone,
Phys. Rev. A \textbf{40}, 3690 (1989).

\bibitem{Bochkarev2026}
M. Bochkarev, N. Solodovchenko, K. Samusev, M. Limonov, T. Wu, and P. Lalanne,
Nanophotonics \textbf{15}, e70071 (2026).

\bibitem{Ott2013}
C. Ott, A. Kaldun, P. Raith, K. Meyer, M. Laux, J. Evers, C. H. Keitel, C. H.
Greene, and T. Pfeifer,
Science \textbf{340}, 716 (2013).

\bibitem{LaneThomas1958}
A. M. Lane and R. G. Thomas,
Rev. Mod. Phys. \textbf{30}, 257 (1958).

\bibitem{PDG2026_Resonances}
D. M. Asner, C. Hanhart, and M. Mikhasenko, ``Resonances,'' Sec.~50 in F.
Takahashi \emph{et~al.} (Particle Data Group),
Int. J. Mod. Phys. A \textbf{41}, 2630011 (2026).

\bibitem{Sekihara2015}
T. Sekihara, T. Hyodo, and D. Jido,
Prog. Theor. Exp. Phys. \textbf{2015}, 063D04 (2015).

\bibitem{HumbletRosenfeld1961}
J. Humblet and L. Rosenfeld,
Nucl. Phys. \textbf{26}, 529 (1961).

\bibitem{Fischbach2026}
J. D. Fischbach, F. Betz, L. Rebholz, P. Garg, K. Frizyuk, F. Binkowski, S.
Burger, M. Hammerschmidt, and C. Rockstuhl,
ACS Photonics \textbf{13}, 5291 (2026).

\bibitem{ShlomoBertsch1975}
S. Shlomo and G. Bertsch,
Nucl. Phys. A \textbf{243}, 507 (1975).

\bibitem{Mizuyama2023_JRPA1}
K. Mizuyama, N. Nhu Le, T. Dieu Thuy, N. Hoang Tung, D. Quang Tam, and T. V.
Nhan Hao,
Phys. Rev. C \textbf{107}, 024303 (2023).

\bibitem{Mizuyama2024_Eigenphase}
K. Mizuyama, T. Dieu Thuy, and T. V. Nhan Hao,
Phys. Rev. C \textbf{109}, 054304 (2024).

\bibitem{Mizuyama2025_JRPA2}
K. Mizuyama and T. Dieu Thuy,
Phys. Rev. C \textbf{111}, 054323 (2025).

\bibitem{Mizuyama2026_Takagi}
K. Mizuyama,
Phys. Rev. C \textbf{114}, 014619 (2026).

\bibitem{Mizuyama2026_Collectivity}
K. Mizuyama,
Phys. Rev. C \textbf{114}, 044326 (2026).

\bibitem{Autonne1915}
L. Autonne,
Ann. Univ. Lyon Nouv. S\'er. I Fasc. \textbf{38}, 1 (1915).

\bibitem{Takagi1924}
T. Takagi,
Jpn. J. Math. \textbf{1}, 83 (1924).

\bibitem{KinugawaHyodo2025}
T. Kinugawa and T. Hyodo,
Eur. Phys. J. A \textbf{61}, 154 (2025).

\bibitem{Co2013}
G. Co', V. De Donno, M. Anguiano, and A. M. Lallena,
Phys. Rev. C \textbf{87}, 034305 (2013).

\bibitem{Hoyer2004}
P. O. Hoyer,
J. Mach. Learn. Res. \textbf{5}, 1457 (2004).

\bibitem{Hurley2009}
N. Hurley and S. Rickard,
IEEE Trans. Inf. Theory \textbf{55}, 4723 (2009).

\bibitem{Burkert2023}
V. Burkert, V. Crede, E. Klempt, K. V. Nikonov, J. A. Oller, J. R. Pel{\'a}ez,
J. Ruiz de Elvira, A. Sarantsev, L. Tiator,
U. Thoma, and R. Workman,
Phys. Lett. B \textbf{844}, 138070 (2023).

\bibitem{Workman2013}
R. L. Workman, L. Tiator, and A. Sarantsev,
Phys. Rev. C \textbf{87}, 068201 (2013).

\bibitem{Feshbach1958}
H. Feshbach,
Ann. Phys. (N.Y.) \textbf{5}, 357 (1958).

\bibitem{Feshbach1962}
H. Feshbach,
Ann. Phys. (N.Y.) \textbf{19}, 287 (1962).

\bibitem{WangMo2020}
J. Wang and Y. Mo,
Phys. Rev. A \textbf{102}, 050803(R) (2020).

\bibitem{KabachnikSazhina1976}
N. M. Kabachnik and I. P. Sazhina,
J. Phys. B \textbf{9}, 1681 (1976).

\bibitem{Gorin2005}
T. Gorin,
J. Phys. A \textbf{38}, 10805 (2005).

\bibitem{Alhassid2006}
Y. Alhassid, Y. V. Fyodorov, T. Gorin, W. Ihra, and B. Mehlig,
Phys. Rev. A \textbf{73}, 042711 (2006).

\bibitem{Clerk2001}
A. A. Clerk, X. Waintal, and P. W. Brouwer,
Phys. Rev. Lett. \textbf{86}, 4636 (2001).

\bibitem{Baernthaler2010}
A. B\"arnthaler, S. Rotter, F. Libisch, J. Burgd\"orfer, S. Gehler, U. Kuhl, and
H.-J. St\"ockmann,
Phys. Rev. Lett. \textbf{105}, 056801 (2010).

\bibitem{Starace1977}
A. F. Starace,
Phys. Rev. A \textbf{16}, 231 (1977).

\bibitem{Ceci2013}
S. Ceci, M. Korolija, and B. Zauner,
Phys. Rev. Lett. \textbf{111}, 112004 (2013).

\bibitem{Shore1967}
B. W. Shore,
J. Opt. Soc. Am. \textbf{57}, 881 (1967).

\bibitem{GorelikUrin2001}
M. L. Gorelik and M. H. Urin,
Phys. Rev. C \textbf{64}, 047301 (2001).

\bibitem{HamamotoSagawa2002}
I. Hamamoto and H. Sagawa,
Phys. Rev. C \textbf{66}, 044315 (2002).

\bibitem{Paliwal2021}
P. Paliwal, A. Blech, C. P. Koch, and E. Narevicius,
Nat. Commun. \textbf{12}, 7249 (2021).

\bibitem{Tu2015}
B. Tu \emph{et~al.},
Phys. Rev. A \textbf{91}, 060502(R) (2015).

\bibitem{MuravievUrin1994}
S. E. Muraviev and M. H. Urin,
Nucl. Phys. \textbf{A572}, 267 (1994).

\bibitem{Urin2008}
M. H. Urin,
Nucl. Phys. \textbf{A811}, 107 (2008).

\bibitem{Watson1952}
K. M. Watson,
Phys. Rev. \textbf{88}, 1163 (1952).

\bibitem{Watson1954}
K. M. Watson,
Phys. Rev. \textbf{95}, 228 (1954).

\bibitem{BohrMottelson1969}
A. Bohr and B. R. Mottelson,
\emph{Nuclear Structure}, Vol.~I (Benjamin, New York, 1969).

\bibitem{Johnson1977}
B. R. Johnson,
J. Chem. Phys. \textbf{67}, 4086 (1977).

\bibitem{Johnson1978}
B. R. Johnson,
J. Chem. Phys. \textbf{69}, 4678 (1978).

\bibitem{Kravanja2000}
P. Kravanja and M. Van Barel,
\emph{Computing the Zeros of Analytic Functions}, Lecture Notes in Mathematics
Vol.~1727 (Springer, Berlin, 2000).

\bibitem{Sakurai2003}
T. Sakurai and H. Sugiura,
J. Comput. Appl. Math. \textbf{159}, 119 (2003).

\bibitem{Fan2003}
S. Fan, W. Suh, and J. D. Joannopoulos,
J. Opt. Soc. Am. A \textbf{20}, 569 (2003).

\end{thebibliography}
\end{document}